\documentclass[sigplan,10pt]{acmart}
\renewcommand\footnotetextcopyrightpermission[1]{}
\AtBeginDocument{%
  \providecommand\BibTeX{{%
    \normalfont B\kern-0.5em{\scshape i\kern-0.25em b}\kern-0.8em\TeX}}}

\setcopyright{acmcopyright}
\copyrightyear{2021}
\acmYear{2021}
\acmDOI{XX.XXXX/XXXXXXX.XXXXXXX}

\acmConference[SOSP '21]{SOSP '21: ACM Symposium on Operating Systems Principles}{Oct. 25-28, 2021}{Germany}
\acmBooktitle{SOSP '21: ACM Symposium on Operating Systems Principles, October 25-28, 2021, Koblenz, Germany}
\acmPrice{15.00}
\acmISBN{978-1-4503-XXXX-X/21/06}

\usepackage{comment}
\usepackage{algorithm}
\usepackage{multirow}
\usepackage{algpseudocode}
\usepackage{subfigure}
\usepackage{setspace}
\usepackage{makecell}
\newcommand{\papertitleabbr}{Lightning}

\newcommand{\victimdevice}{GPU}

\begin{document}
\title{Lightning: Striking the Secure Isolation on GPU Clouds with Transient Hardware Faults}

\author{Pengfei Qiu}
\authornotemark[1]
\email{qpf15@tsinghua.org.cn}
\affiliation{%
  \institution{Tsinghua University}
  \city{Beijing}
  \country{China}
}

\author{Rihui Sun}
\authornote{Both authors contributed equally to this research.}
\author{Jian Dong}
\email{19b903009@stu.hit.edu.cn}
\email{dan@hit.edu.cn}
\affiliation{%
  \institution{Harbin Institute of Technology}
  \city{Harbin}
  \country{China}
}

\author{Yongqiang Lyu}
\authornote{Corresponding author.}
\author{Haixia Wang}
\email{{luyq, hx-wang}@tsinghua.edu.cn}
\affiliation{%
  \institution{Tsinghua University}
  \city{Beijing}
  \country{China}
}

\author{Ningxuan Feng}
\author{Peichen Guo}
\email{{fnx20,gpc18}@mails.tsinghua.edu.cn}
\affiliation{%
  \institution{Tsinghua University}
  \city{Beijing}
  \country{China}
}

\author{Gang Qu}
\email{gangqu@umd.edu}
\affiliation{%
  \institution{University of Maryland}
  \city{College Park, Maryland}
  \country{USA}
}

\author{Dongsheng Wang}
\email{wds@tsinghua.edu.cn}
\affiliation{%
  \institution{Tsinghua University}
  \city{Beijing}
  \country{China}
}

%\title{Exposing the Threat of Hardware Voltage and Frequency Faults on Shared GPU Clouds: To Non-Intrusively Make DNNs Untrusted}
%\title{Thunder: Making DNNs Untrusted on Shared GPU Clouds by Transient Hardware Voltage and Frequency Faults}

\begin{abstract}
GPU clouds have become a popular computing platform because of the cost of owning and maintaining high-performance computing clusters. Many cloud architectures have also been proposed to ensure a secure execution environment for guest applications by enforcing strong security policies to isolate the untrusted hypervisor from the guest virtual machines (VMs). 
%However, these secure isolation still lack of concern about potential vulnerability induced by untrusted GPU hardware. 
%In this paper, we propose the Lightning to show a potential security threat caused by transient hardware faults generated by exploiting the Dynamic Voltage and Frequency Scaling (DVFS) technology, which can mislead the Deep Neural Network (DNN) inference inside the GPU cloud execution environment with secure isolation.
In this paper, we study the impact of GPU chip's hardware faults on the security of cloud "trusted" execution environment using Deep Neural Network (DNN) as the underlying application. We show that transient hardware faults of GPUs can be generated by exploiting the Dynamic Voltage and Frequency Scaling (DVFS) technology, and these faults may cause computation errors, but they have limited impact on the inference accuracy of DNN due to the robustness and fault-tolerant nature of well-developed DNN models.
  
To take full advantage of these transient hardware faults, we propose the \papertitleabbr\ attack to locate the fault injection targets of DNNs and to control the fault injection precision in terms of timing and position. %More specifically, we design a sensitive targets search algorithm to find the most critical processing units in the DNN models that determine the inference results. Then we use a genetic algorithm to automatically optimize the fault injection parameters. 
We conduct experiments on three commodity GPUs to attack four widely-used DNNs. Experimental results show that the proposed attack can reduce the inference accuracy of the models by as high as 78.3\% and 64.5\% on average. More importantly, 67.9\% of the targeted attacks have successfully misled the models to give our desired incorrect inference result. This demonstrates that the secure isolation on GPU clouds is vulnerable against transient hardware faults and the computation results may not be trusted. %In addition to the DNNs, we suggest that the attack may also break the security of the encryption methods such as AES and RSA inside the secure isolation, which presents a key issue to the future secure isolation architecture design for both cloud and GPU.
\end{abstract}



\settopmatter{printfolios=true}

\maketitle

\section{Introduction}
%With the development of the Deep Neural Network (DNN) technology, the model structure is becoming more and more complex and the number of the parameters is increasing rapidly. The computing resources for training and performing the complex DNN models is massive.
%With the development of the computing technology, especially the Deep Neural Network (DNN) whose model structure is becoming more and more complex and the number of the parameters is increasing rapidly, 
%With the development of the advanced medical imaging for battling cancer, automated customer service, cinematic-quality gaming, next-generation capabilities in AI, high-performance computing (HPC), the computing resources required for those cumbersome tasks are massive. Using the individual devices to complete these cumbersome tasks is therefore time-consuming and uneconomical. In order to meet the requirement for the high computing resources and low-cost, 
Leading cloud providers such as Amazon, Google, IBM, Baidu, and Alibaba offer their users the access to Graphics Processing Unit (GPU) computing resources as transparently available computing power or as a part of leased virtual machines (VMs) in a multi-tenant manner, which is referred to as GPU cloud~\cite{lombardi2014towards, sengupta2013multi, NvidiaCloud}, in which users can fulfil their Deep Neural Network (DNN) training or inference services just in the simple pay-as-you-go ~\cite{qi2014vgris}. 
%xu2021aspdac. Security of Neural Networks from Hardware Perspective: A Survey and Beyond 

In current cloud computing paradigm, the cloud hypervisors have been eventually recognized as untrusted by the guest users due to the security and privacy concerns on the two kinds of situations, i.e., an untrusted cloud vendor having full access to the guest contents by owning the hypervisor privilege, or a malicious guest user who is able to gain the hypervisor's privilege by utilizing the undisclosed vulnerabilities in the cloud software~\cite{mi2020mostly}. % Even if the former factor might be mitigated by a good business model or trust chain, the latter is hardly possible to be resolved due to the high complexity of cloud software and cost of sufficient formal verification on it. 

In order to protect the guest virtual machines especially the valuable intellectual properties of the DNN models as well as the valuable and sensitive user data~\cite{xu2021aspdac} from the potential security or privacy compromise induced by the untrusted hypervisor situations, a range of secure cloud architectures in the domain may be deployed,%by enforcing strong security policies to isolate the untrusted hypervisor from the guest virtual machines (VMs), 
such as CloudVisor~\cite{zhang2011cloudvisor}, CloudVisor-D~\cite{mi2020mostly}, H-SVM~\cite{jin2011architectural}, HyperWall~\cite{szefer2012architectural}, NOVA~\cite{steinberg2010nova}, Xoar~\cite{colp2011breaking}, Nexen~\cite{shi2017deconstructing}, HyperLock~\cite{wang2012isolating}, DeHype~\cite{wu2013taming}, and HypSec~\cite{li2019protecting}. The main idea of those methods is to ensure a secure execution environment for guest applications and restrict the privilege of the hypervisor VM monitors from direct access to guest VM memory. 
Similar ideas have been applied to improve the protection of GPU applications by introducing isolated DNN execution environment such as Graviton~\cite{volos2018graviton} and Heterogeneous Isolated eXecution (HIX) ~\cite{Insu2019Heterogeneous}. 

%Hua2003dac Energy reduction techniques for multimedia applications with tolerance to deadline misses

However, the existing defenses against untrusted hypervisors are based on isolation and logic access control, and the potential vulnerabilities induced by the underlying hardware (i.e. GPU chips) faults have not been studied thoroughly. The lack of such study is due to two main reasons. First, hardware faults in modern chips are rare. Second, many applications such as digital signal processing and DNN models are fault-tolerant by design ~\cite{Hua2003dac, li2014scaling}. We argue that when hardware faults are generated by a malicious attacker, they will not be rare. However, whether any effective attacks can be built on such faults is still unknown. In this paper, we answer this question by proposing the Lightning attack, where we create transient hardware faults deliberately during the program's secure isolated execution on GPU clouds in order to cause execution failure.  

%In this paper, we study a GPU hardware fault-based security risk that may break the protection of secure isolation and compromise the trustworthiness of DNNs running inside, which is induced by the Dynamic Voltage and Frequency Scaling (DVFS) technology currently widely used in GPU chips. 

%In the GPU clouds, each single GPU is shared across different customers (i.e. GPU-as-a-Service)~\cite{lombardi2014towards}. Therefore, the design defects in the GPU chips will largely threaten the security of the DNNs preformed on the GPU clouds.  

%Hardware vulnerabilities related to the Dynamic Voltage and Frequency Scaling (DVFS) technology have greatly raised the research attention recently. The previously-disclosed attacks, such as overclocking attack by manipulating the frequency on ARM CPUs~\cite{tang2017clkscrew}, the voltage fault-based attacks to Intel~\cite{qiu2019voltjockeySGX, murdock2020plundervolt, kenjar2020v0ltpwn} and ARM~\cite{qiu2019voltjockey} CPUs, showed that such vulnerable defects of the large-scale integrated circuits can cause severe malfunctions of CPU microarchitectures. Sabbagh et al. also showed that the voltage faults can attack the AES encryptions running on AMD GPU~\cite{Majid2020Novel}, which still lacks further investigation and verification on its effects on DNNs. In emerging computing technologies, GPUs are actually playing a much more important role in accelerating DNNs than encryption and decryption. Investigation of the voltage and frequency-related vulnerabilities on GPU accelerators for DNNs is therefore very urged and valuable. 

In Lightning, the hardware faults are induced by the Dynamic Voltage and Frequency Scaling (DVFS) technology that is available in all commercial GPU chips.
DVFS dynamically adjusts the voltage and frequency of GPUs based on the real-time workloads to balance low power and high performance. For example, on Nvidia GPUs, this is known as the GPU boost technology.
In general, low voltage and/or low frequency results in low power consumption and long execution time. Voltage and frequency are matched in order to satisfy the circuit's timing constraint. Hardware faults may occur when an inappropriately low voltage or a high frequency is used. These faults are in the form of bit flips and last shortly. To investigate whether such faults can cause security breaches, we consider convolutional neural network (CNN) models and validate the fault's impact on the model accuracy, in particular, whether Lightning attack can lead the models to give incorrect results as the attacker's desire.

In sum, we make the following contributions in this paper.

\begin{itemize}
\item We reveal a potential security threat that may break the protection of the secure isolation on GPU clouds by exploiting transient voltage and frequency faults. 
\item We propose the prototype attack method, the Lightning, which can non-intrusively and dramatically mislead DNN inference running inside the secure isolation without any adversarial samples assist. 
%\item In order to improve the efficiency of the exploitation, we propose a sensitive targets search algorithm to locate the best fault injection targets in DNNs, and also propose a genetic algorithm to automatically optimize the faulting parameters for precise fault injections.
%During the attack, the attacker procedure that is executed on CPU will break \importantpartname\ by inducing low-voltage hardware faults and therefore mislead the output of victim \victimobjectabbr\ model that is performed on \victimdevice.
\item We implement and verify the method on real hardware by using three commodity Nvidia GPUs. The experimental results show that the attack can decrease the accuracy of different convolutional neural networks on MNIST, CIFAR-10, and Yale face data sets by $64.5\%$ on average, and can targeted-attack the Lenet-5 model with MNIST data set to misclassify the results by $67.9\%$ on average.
\item In addition to the DNNs, we suggest that the potential threat may also break the security of the encryption methods such as AES and RSA inside the secure isolation, which presents a key issue to the secure isolation architecture design for both clouds and GPUs.   
\end{itemize}

%The remainder of this paper is organized as follows. Section \ref{RelatedWork} introduces the related work including attacks to DNNs, hardware faults-based vulnerabilities, and GPU vulnerabilities. Section \ref{ProposedMethod} describes the discovered security risks and the corresponding exploit. Section \ref{SearchAlgorithm} gives the details of the sensitive target search algorithm for positioning faulting targets for DNNs, and Section \ref{sec:ParamterFind} presents the genetic algorithm for fault injection parameter search, after which, Section \ref{Experiments} shows the verification for the security risk and the attack on the Nvidia GPUs based on those algorithms, and at the end of the paper, the discussion and conclusion of the study are presented in Section \ref{Dissicution} and Section \ref{Conclude}, respectively.

\section{Lightning: GPU Cloud Attack with Transient Hardware Faults}
\label{ProposedMethod}
This study discloses a security risk related to the dynamic voltage and frequency scaling technology (DVFS) in the power subsystem of GPU chips, which is able to attack the DNNs running inside the secure isolation environment on GPU clouds. The entire attack does not rely on any software vulnerabilities.

\subsection{Threat Model}
\label{sub:threatmodel}
\begin{figure}[htb]
\begin{center}
\includegraphics[width=0.9\columnwidth]{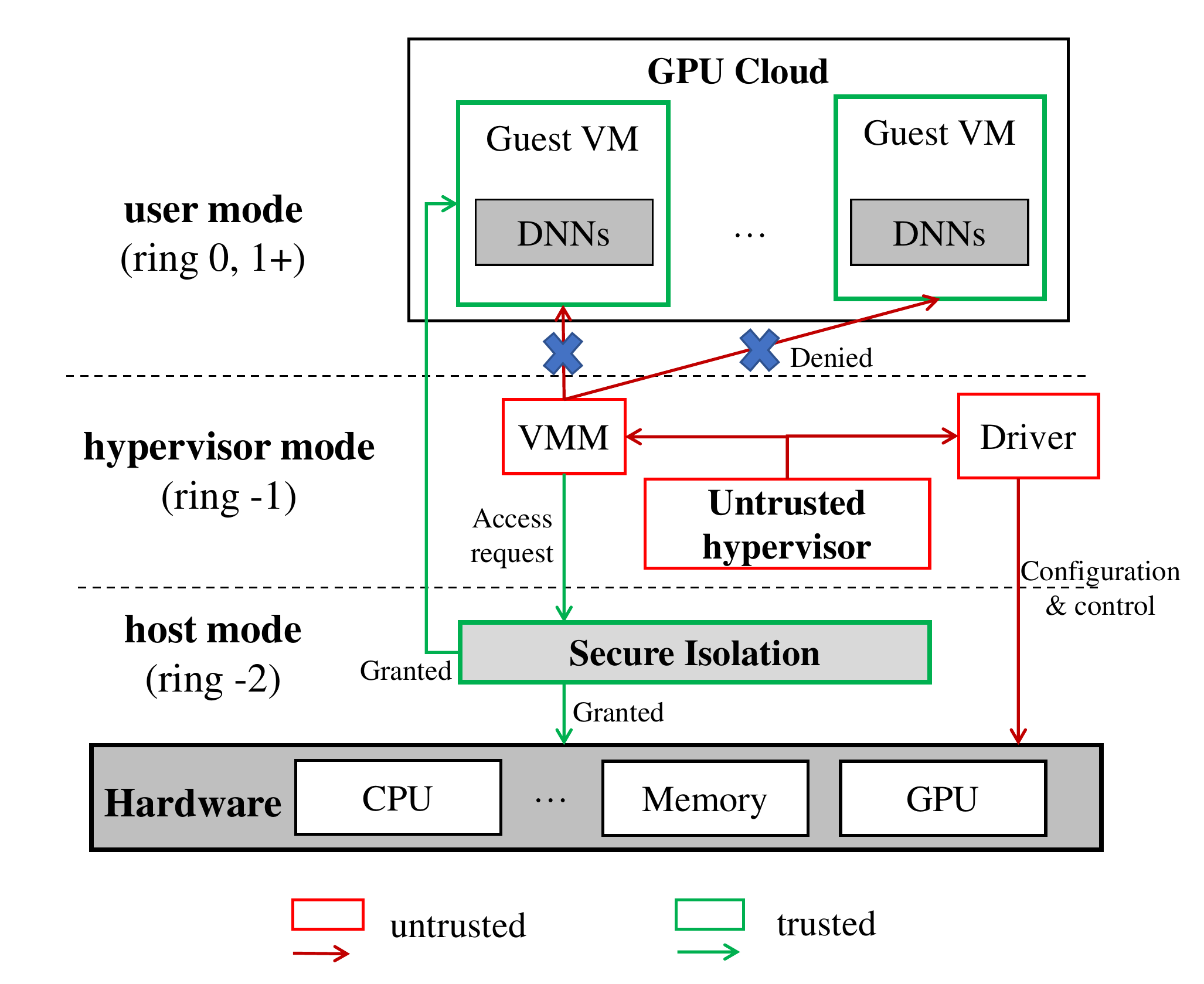}
\caption{The threat model for GPU clouds. This study suggests that the secure isolation for GPU clouds may be bypassed by exploiting the hardware-related vulnerabilities.}
\label{fig:threatmodel}
\end{center}
\end{figure}

This study sticks to the similar threat model as that employed by most of the state-of-the-art secure isolation designs, such as CloudVisor~\cite{zhang2011cloudvisor} and H-SVM~\cite{zhang2011cloudvisor}. The major difference from those works is that a further revision to the assumed trusted hardware is made, i.e., in this study, we assume that the hardware is trusted but not fully fault-tolerant. Figure \ref{fig:threatmodel} shows the details.

%On the victim side, the voltage and frequency of the target GPU accelerators should be able to be manipulated using software instructions. This is valid for most of the current GPU chips in which frequency and voltage scaling technology is enabled.

%On the attacker side, this study assumes the hypervisor is untrusted as many other cloud security work does~\cite{zhang2011cloudvisor, mi2020mostly, jin2011architectural, szefer2012architectural, steinberg2010nova, colp2011breaking, shi2017deconstructing,wang2012isolating, wu2013taming, li2019protecting}, and the hypervisor has been defended against in free access to the user data in virtual machines, including those in GPU's memories, except that the root can still control the software drivers for the voltage and frequency manipulation of GPU devices. The key significance of Thunder attack in such scenarios should be noticed that the attackers in this scenario cannot modify or replace the model weights or structures of the targeted DNNs running on GPUs even if they have acquired a root privilege.
%This study assumes the hypervisor is untrusted as many other cloud security work does~\cite{zhang2011cloudvisor, mi2020mostly, jin2011architectural, szefer2012architectural, steinberg2010nova, colp2011breaking, shi2017deconstructing,wang2012isolating, wu2013taming, li2019protecting}. 
In order to protect guest VM's security and privacy, the state-of-the-art secure cloud designs propose to build a secure isolation between the untrusted hypervisor and the guest VMs to prevent the guest VM's memory (including the GPU memory) from being directly accessed by the untrusted hypervisor via Virtual Memory Monitor (VMM). %However, the hypervisor can still control the devices via the drivers since they are in the same privilege ring and are regarded as the same untrusted. 
The detailed assumptions of the threat model are listed as follows.
\begin{itemize}
    \item A secure isolation between the untrusted hypervisor and the guest VMs is build, and the direct access to guest VM's memory as well as the GPU memory from the hypervisor is denied.
    \item The hypervisor can still control the drivers to the underlying hardware such as CPU, memory and GPUs for task scheduling or power control. 
    \item Guest DNNs are running on the guest VMs and are accelerated by GPUs assigned to the guest VMs. 
  \end{itemize} 
We propose the \papertitleabbr\ to verify that exploiting transient voltage and frequency faults generated by DVFS can bypass the secure isolation and attack the DNNs on GPU clouds directly, which are actually not considered in the state-of-the-art secure architecture designs. In addition to the previous findings on the security impact of the DVFS faults on AES and RSA encryption within secure isolations~\cite{qiu2019voltjockey, qiu2019voltjockeySGX}, we strongly argue that a considerate design methodology against both the logical access vulnerabilities and the physical hardware vulnerabilities should be taken into account for better cloud and GPU securities.

\subsection{Security Risks in the Voltage and Frequency Scaling Technology of GPUs}
\label{sub:PowerSubsystem}
\begin{figure}[htb]
\begin{center}
\includegraphics[width=0.85\columnwidth]{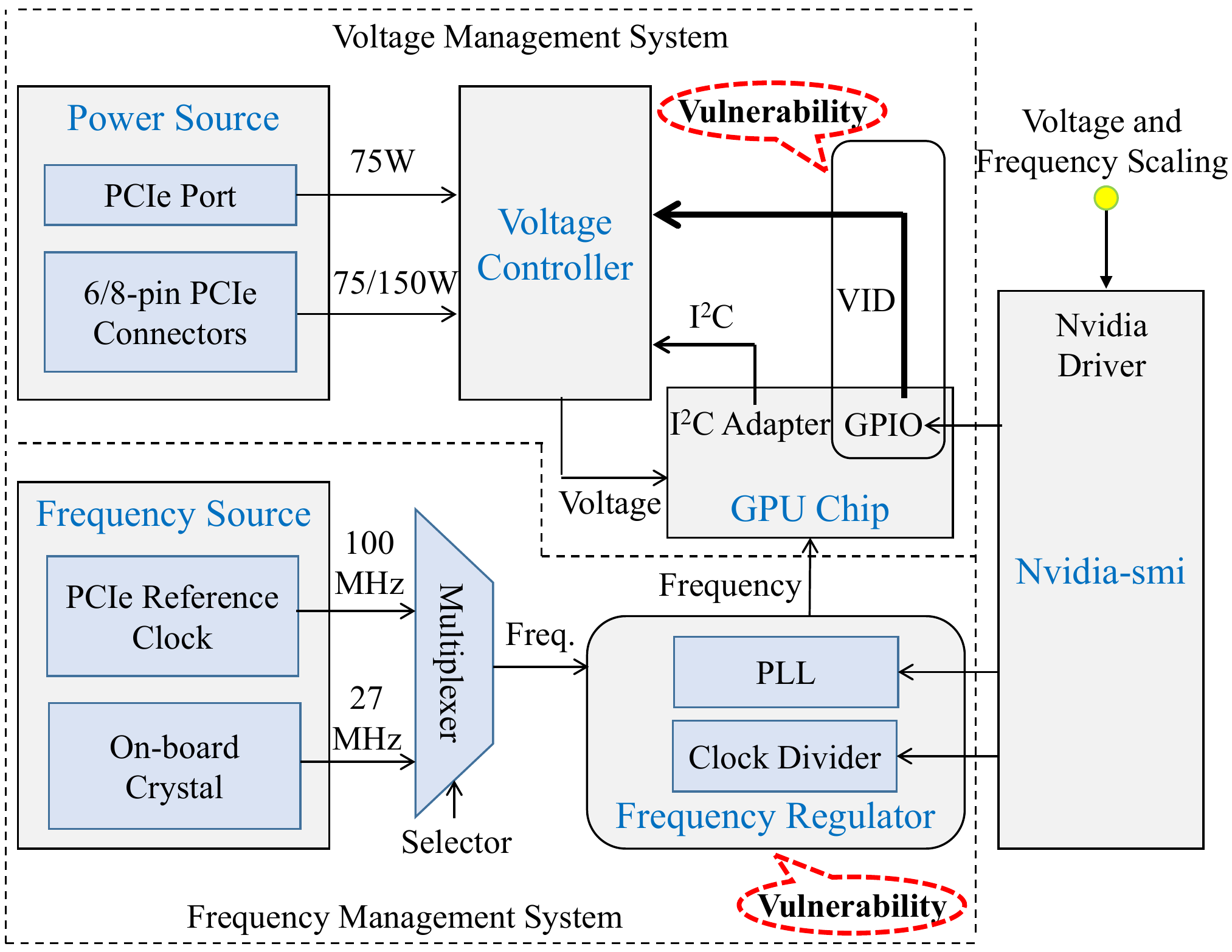}
\caption{The voltage and frequency management framework of Nvidia {\victimdevice}s. Both the voltage and frequency can be adjusted by software drivers.}
\label{fig:powersubsytem}
\end{center}
\end{figure}

The voltage and frequency are significant factors to influence the dynamic power\footnote{Formally, the dynamic power $P_{t}$ of a circuit at time t is: $P_{t} \propto V_{t}^{2}F_{t}C$, where $V_{t}$ is the voltage, $F_{t}$ is the frequency, and C is the capacitative load.} of an integrated circuit~\cite{hong1999power}, and current \victimdevice\ accelerators have widely adopted the voltage and frequency scaling technology in their power subsystems to manage the power consumption while maintaining necessary performance according to real-time workloads. Using Nvidia GPU design as an example, its voltage and frequency management framework is illustrated in Figure \ref{fig:powersubsytem}, where there are security risks disclosed in this study. 

Nvidia GPUs are usually powered by the PCIe port or PCIe connectors. The PCIe port and 6-pin PCIe connector can drive GPU with the power ups to 75W, and the 8-pin PCIe connector can give GPU a power ups to 150W. The PCIe and PCIe connectors can provide different voltages that are higher than the normal operating voltage. The voltage controller is responsible for the DC-DC conversion (voltage conversion) which sources from the PCIe port or PCIe connectors and outputs appropriate voltages to GPU. The voltage output of voltage controller is software-controllable through the General-Purpose Input/Output (GPIO) pin (or the $I^{2}C$ interface managed by the $I^{2}C$ adapter on some high-end GPU cards), which is governed by the Nvidia System Management Interface (\emph{Nvidia-smi} driver). It permits administrators to modify and query the device states of GPUs, and thus becomes one of the root causes of the potential defect disclosed in this study. We verify that the working voltages of Nvidia GPUs can be lowered to a very low value with software instructions performed on CPU via the \emph{Nvidia-smi} drivers, which enables the attackers to generate transient voltage glitches to achieve hardware fault injection attacks on DNN models running on Nvidia GPU accelerators. 

Furthermore, the potential security risks also exist in the frequency management system of Nvidia GPUs. There are two clock sources on the current GPU chips, i.e., the PCIe reference clock (100 MHz) and an on-board crystal (usually 27 MHz)~\cite{peres2013reverse}. A multiplexer is employed to select the clock source, and the system frequency can be increased by the Phase-Locked Loop (PLL) or lowered by the clock divider whose working parameters can be configured through the \emph{Nvidia-smi} driver. We verify that the Nvidia GPUs do not limit the outputs of PLL or clock divider and can be easily overclocked by setting the frequency offset with \emph{Nvidia-smi}, which also potentially enables the frequency-based hardware fault injection attacks.

In addition to the undervoltaging or overclocking situations, it should be noticed that the adjustment to the voltage and frequency for the attack does not need to be outside of the legal ranges. Hardware using DVFS technologies may provide a legal voltage-frequency table for the software monitor to choose appropriate voltage-frequency pairs from. However, when such a pairing scheme for specific voltage vs. frequency is not hardware-guaranteed, i.e., the legal values of voltages and frequencies in the given range can be freely paired, such a mismatch can still induce hardware faults. For example, for a given working frequency, a system-recommended paired voltage is normally enough to maintain it, but a timing violation-related hardware fault may occur when the attacker tunes to a lower voltage (in the legal range) that actually cannot sufficiently support the given working frequency. In practice, the attackers can often choose a high but legal frequency for the GPU and choose a low but legal voltage that is insufficient for supporting the frequency, which is helpful for a stealthy attack.

\subsection{Exploitation to Attack DNNs}
\label{sub:Exploration}
In order to exploit the voltage and frequency-related feature aforementioned in Section \ref{sub:PowerSubsystem} to attack the victim DNN model that is accelerated by a GPU chip, an attacker procedure that is firstly launched on CPU, as Figure \ref{fig:Overview} demonstrates, can generate transient low-voltage or high-frequency glitches to the GPU at dedicated timings to induce hardware faults, by which the execution of the victim DNN model may be compromised.

Compared to DNN training, DNN inference is relatively easy and fast to be prototype verified for attack yet widely deployed in current GPU cloud-based business, such as for mobile devices to reduce lower time and energy consumption~\cite{TGuo2017cloudbased}, direct model serving systems~\cite{Cloudlets} to reduce prediction latency and improves
throughput, or cloud vendor such as Nvidia GPU Cloud to provide pre-trained models for inference services~\cite{NvidiaNGC}. This study aims at the verification of attacking DNN inferences on GPU clouds.
%The hardware faults happening on the GPUs will have no effect on the attacker procedures running on CPU, which makes the attack much easier to gain a better success rate than the conventional CPU-based fault injection attacks for ARM TrustZone or Intel SGX~\cite{qiu2019voltjockey, qiu2019voltjockeySGX}. 

\subsubsection{Victim DNN Model}
\begin{figure}[htb]
\begin{center}
\includegraphics[width=0.98\columnwidth]{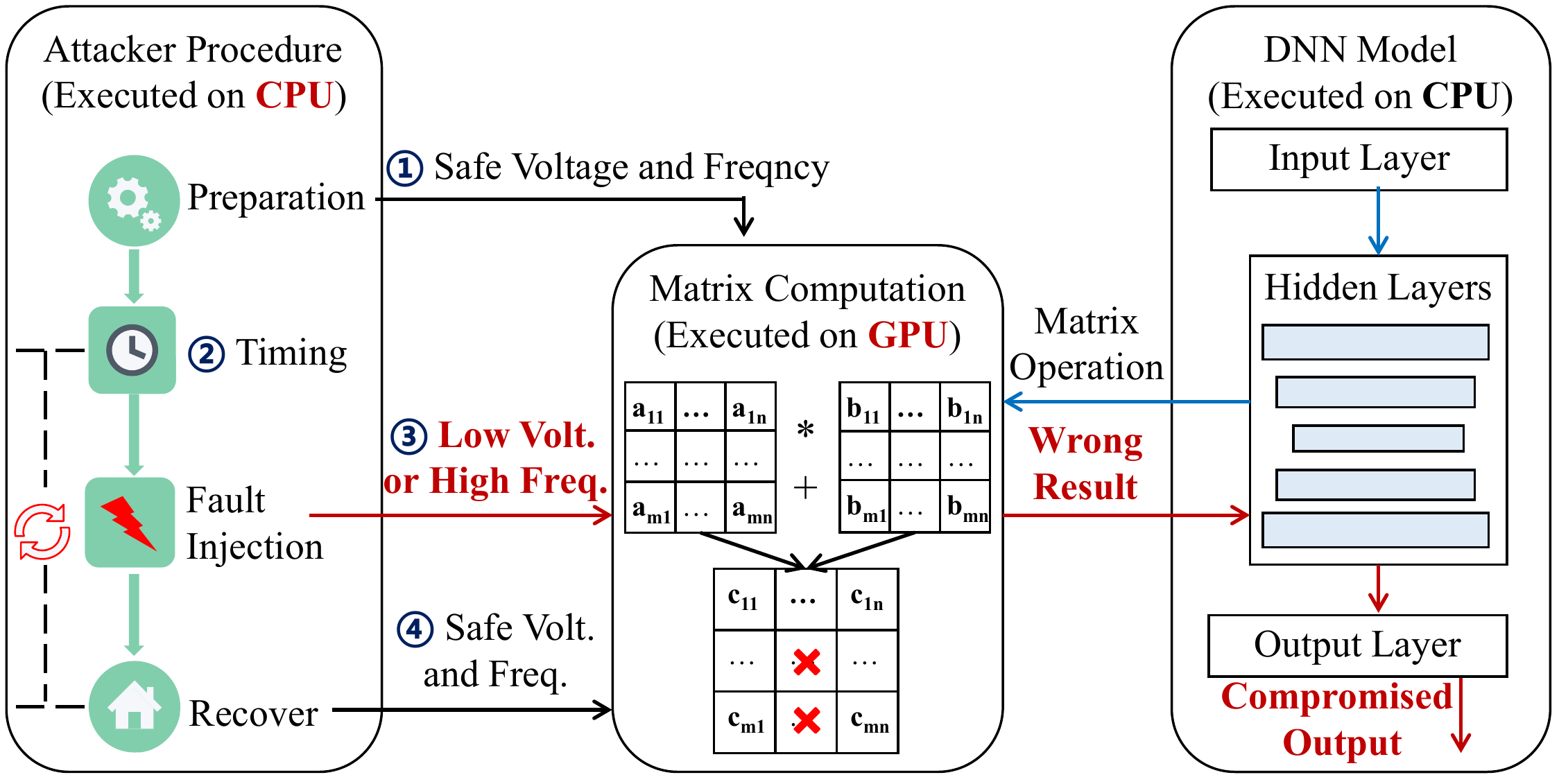}
\caption{The exploitation flow for DNN model attack. The attacker procedure takes four steps to complete the attack: \textcircled{1} configuring CPU and GPU with a safe voltage and frequency; \textcircled{2} waiting for fault injection points; \textcircled{3} creating low-voltage or high-frequency glitches to induce faults into GPU; \textcircled{4} recovering the voltage and frequency of GPU.}
\label{fig:Overview}
\end{center}
\end{figure}

The right side of Figure \ref{fig:Overview} illustrates the execution flow of the victim DNN model. Note that the victim DNN model is performed in the guest VMs of the GPU clouds and the attacker has no way to touch the weights and model structures of the target DNN model. The CPU first loads the DNN model and its corresponding parameters from the corresponding guest VM' private memory, and then feeds the input to the DNN model. Next, the layers of the DNN model is successively executed to identify the input. In order to accelerate the inference process, the computation-intensive matrix operations such as convolutions in Convolutional Neural Networks (CNNs) are usually configured to be scheduled from CPU to GPU, which actually plays a very important role in determining the model classification result. Consequently, influencing the matrix operations by injecting hardware faults may compromise the DNN model's outputs. 
\subsubsection{Attacker Procedure}
\label{subsec:overattacker}
As Figure \ref{fig:Overview} shows, there are four standard attacker routines in the attacker procedure, which are described as follows.

\emph{\bf{Preparation.}} Before conducting fault injections, preparing a suitable attack baseline is helpful to improve the attack performance. Firstly, the attacker procedure sets the GPU and CPU with a safe voltage and frequency.%, in which the CPU's voltage and frequency will be kept unchanged throughout the entire attack processes in order to reduce the side effects of timing changes caused by the DVFS of CPU. Different from CPU, the voltages and frequencies of the GPU are just kept until the execution flow of victim DNN system reaches the fault injection points where hardware faults will be injected. 
Secondly, all the residual states of the GPU should be cleared because these states may affect the execution speed of the victim DNN system and thus impact the timing calculation for the fault injection to start, for example, cache layouts, branch prediction table, interrupt vector table and status registers, etc. 

\emph{\bf{Timing.}} The attack goal is achieved by inducing hardware faults at expected timing points of DNN inference execution. %Once the victim DNN system is launched, the attacker procedure starts to count clock cycles to evaluate the execution of the victim DNN system. 
As soon as the execution flow arrives at the expected target locations to inject faults, the attacker procedure will create a hardware fault to influence the execution of the victim DNN inference. Such a timing for waiting for the target to be injected with faults is critical for the success rate of the attack.%, and the algorithms for determining it are presented in Section \ref{SearchAlgorithm} and Section \ref{sec:ParamterFind}.

\emph{\bf{Fault injection.}} This step is responsible for triggering hardware faults. The value of glitch voltage-frequency pair and its lasting duration are two key factors to induce controllable and effective faults. The induced hardware faults in GPU circuit last too long may cause the hardware crash or no response. Therefore, it is very important to find out a proper voltage-frequency pair and its lasting duration that will result in desired data changes.% We utilize a genetic algorithm to search appropriate fault injection parameters for the best performance, which is presented in Section \ref{sec:ParamterFind}.

\emph{\bf{Recover.}} After a fault is successfully injected at the expected target, the GPU's voltage and frequency should be recovered to the baseline to resume the victim DNN inference normally, which is then ready for the next cycle of fault injection staring with the waiting step.

\subsubsection{Fault Injection Parameters}
To achieve the \papertitleabbr\ attack, the following parameters must be firstly determined, i.e.,

$P_{fault} = \{F_{C}; V_{C}; F_{G}; V_{G}; F_{h}; V_{l}; T_{W}; T_d\}$,

whose definitions are shown in Table \ref{table:ParameterMeaning} respectively.

\begin{table}[htbp]
\caption{Fault injection parameters of \papertitleabbr\ attack.}
\centering
\begin{spacing}{1.19}
\begin{tabular}{p{0.03\columnwidth}p{0.9\columnwidth}}
\hline
\small{Para.} & \makecell[c]{\small{Meaning}} \\
\hline
\multirow{1}{*}{\small{$F_\mathit{C}$}} & \small{The frequency for CPU, which is unchanged during the attack.}\\
\multirow{1}{*}{\small{$V_\mathit{C}$}} & \small{The voltage for CPU, which is unchanged during the attack.} \\
\multirow{1}{*}{\small{$F_\mathit{G}$}} & \small{The baseline and restore frequency for \victimdevice, which is a fixed value.} \\
\multirow{2}{*}{\small{$V_\mathit{G}$}} & \small{The baseline and restore voltage for \victimdevice, which is a fixed value and should be sufficient to maintain $F_G$.} \\
\multirow{1}{*}{\small{$F_\mathit{h}$}} & \small{The glitch frequency at which faults can be induced to \victimdevice.} \\
\multirow{2}{*}{\small{$V_\mathit{l}$}} & \small{The glitch voltage at which faults can be induced to \victimdevice\ (high voltages may damage circuits and are not used in this study).} \\
\multirow{2}{*}{\small{$T_\mathit{W}$}} & \small{The timing for the attacker procedure to wait for to induce hardware faults at target locations.} \\
\multirow{2}{*}{\small{$T_\mathit{d}$}} & \small{The period that $F_{h}$ or $V_l$ should be hold for to induce the expected hardware faults.} \\
\hline
\end{tabular}
\end{spacing}
\label{table:ParameterMeaning}
\end{table}

\subsubsection{Challenge of Exploitation}
\label{subsub:Challenge}
Intuitive attack ideas can hardly work in practice because randomly induced faults have very limited impact on the inference accuracy of DNNs due to the robustness and fault-tolerant nature of well-developed DNN models. A random fault injection experiment with uncontrolled strategies is conducted for four CNN models on Nvidia GeForce GTX 1650 GPU, as shown in Table \ref{table:Randomtest}. It shows that the model inference accuracy degradation is very limited. So, a major challenge of the proposed attack is how to improve the efficiency for taking full advantage of the hardware faults. To achieve this, we propose a fast gradient-based sensitive target searching algorithm to help position the most critical processing units in the DNN models on GPUs that impact the inference results mostly, based on which a genetic faulting parameter searching algorithm is proposed to automatically optimize the fault injections for the goal.

\begin{table}[htbp]
\caption{Random fault injection on Nvidia GeForce GTX 1650 GPU}
\centering
\small
\begin{tabular}{p{0.2\columnwidth}|p{0.2\columnwidth}|p{0.2\columnwidth}|p{0.2\columnwidth}}
\hline
\multirow{1}{*}{Model} & \multirow{1}{*}{Data } & \multirow{1}{*}{Base} & \multirow{1}{*}{Accuracy}\\
 & Set & accuracy & degadation\\
\hline
Lenet-5 & MNIST & 98.2\% & 4.2\%\\

\hline
AlexNet & MNIST & 98.7\% & 1.9\%\\
\hline

ResNet-18 & CIFAR-10 & 97.6\% & 3.5\%\\
\hline

MobileNet & Yale & 98.8\% & 6.8\%\\
\hline
\end{tabular}
\label{table:Randomtest}
\end{table}
%Because the state-of-the-art DNNs are already fault-tolerant, the ramdom induced faults in DNN models

\section{Fast Gradient-Based Sensitive Targets Search}
\label{SearchAlgorithm}
%An important challenge in using fault injections for efficient DNN attacks is to find out the sensitive targets of the DNN model where the faults can result in the most effective disturbance on the final output. 
%As shown in Figure \ref{fig:faultlocation}, GPU is scheduled by CPU to execute computation-intensive operations for DNNs, such as matrix multiplication in convolutional layers for CNN. When there are faults injected into the computing process, its faulty results will be accepted without any error corrections into the following data structures and processes, such as feature maps and subsequent neural networks, which may lead to a wrong classification result output. 
The main reason why the transient hardware faults can compromise a DNN is due to the model reliability issues with respect to the variations of data and in-process signals in the current neural network-based cognition models, which is also widely exploited by adversarial example attacks, such as FGSM attack~\cite{goodfellow2014explaining}. In this study, deep Convolutional Neural Networks (CNNs) are used for the verification of the \papertitleabbr\ attack, and a sensitive target positioning algorithm based on gradient search is proposed to help find out the most sensitive targets of a given CNN model, on which the incoming faults may result in the most effective disturbance to the final outputs. The algorithm adopts a similar idea with FGSM attack.%, and assumes that the model structure of target model to be attacked is knownby the attacker.
%\begin{figure}[htb]
%\begin{center}
%\includegraphics[width=0.85\columnwidth]{pic/modelprocess.pdf}
%\caption{Fault injection-based attack on a GPU-accelerated CNN. The hardware faults influence the calculation process of convolution and result in bit flips in the feature maps, which will be transferred to the input of the next layer and then be propagated eventually to the final output.}
%\label{fig:faultlocation}
%\end{center}
%\end{figure}

\subsection{Problem Deﬁnition}

%Random fault injections to a \victimobjectabbr\ cannot degrade the model output accuracy notably, and the time consuming for the attack is also very high. Pilot testing by such methods on Lenet-5 model with MNIST data set on the Nvidia GeForce MX230, GeForce 1050, and GeForce 1650 GPUs can only reduce the output accuracy by $0.2\%$ on average after overwhelming trials. In addition, multiple faults may impact the inference results more effectively than single fault, but multiple faults often need longer total hold periods ($\sum\limits_{i=1}^{N}{T_{d}}$, where N is the number of induced hardware faults) for abnormal voltage and frequency glitches, which means a lower success rate due to the high risk of hardware crash or halt. 
%In this study, a sensitive target search algorithm is proposed to resolve the challenges in Section \ref{sub:Exploration}, which mainly sticks to the single bit-flip faults and can help the attacker procedure to lower the model output accuracy to the maximum extent with the minimum injected faults.

Taking CNN as the target for verification, given a CNN that contains $L$ layers, the input of the network is the vector $x$, and the output vector is $t$. The feature maps over the layer from $1$ to $L$ that the inference computation generates for $x$ can be referred to as a feature map set, in which each feature map consists of a certain number of elements (e.g. $m$ floating-point numbers) computed by the corresponding convolution operations between the last-layer feature maps and convolution kernels. The elements of all the feature maps over all the layers can be referred to a set $E$, each of which will be presented as an arithmetic word (with length $w$) according to the hardware specifications of GPU, e.g., a 4-byte floating-point number and therefore can be expressed as a machine bit vector $[b_{1}, b_{2}, ..., b_{w}]$ as defined in Equation (\ref{equ:ElementLayer}).

\begin{equation}
\begin{aligned}
\label{equ:ElementLayer}
& \ E = \{e_{k}^{j} \mid e_{k}^{j} = [b_{1}, b_{2}, ..., b_{w}], \\
& \qquad \text{$j$ for each feature map in all layers}, \\
& \qquad  \text{$k$ for each element in the feature map}, \\
& \qquad \text{$w$ is the arithmetic word width of an element} \ \ \}
\end{aligned}
\end{equation}

Our goal is to find out the most sensitive elements from $E$ whose bit flips can induce the most significant network accuracy degradation. We leverage the original loss function employed during training the CNN model as the metric to evaluate the influence of the fault injections on given elements, the optimization goal of which in this study is to increase the loss as much as possible and consequently to make the output vector cross the classification boundary. We regard the elements whose bit flips can result in larger loss increase as more sensitive ones. As mentioned above, injecting multiple hardware faults may be helpful to increase the loss, but it may crash the GPU. Therefore, in this study, we constrain the voltage and frequency faults-induced bit flips to only \textit{1} bit for each sensitive target to keep the high performance and success rate of the fault injections. The sensitive target search problem can be formulated as the optimization problem shown in Equation (\ref{equ:Problem}), which aims to use limited hardware faults to gain maximal loss.

\begin{equation}
\begin{aligned}
\label{equ:Problem}
&max \ loss(f(x; E_{N}),t) \\
&s.t.\ E_{N} \subseteq E, \mid{E_{N}} \mid \leq N \\
%where \ B^{l} = { B_{i}^{l} | i = 1, ..., m}, and B{i}=[b_{1}, b_{2}, ..., b_{w}]
\end{aligned}
\end{equation}

The $f$ represents the trained victim CNN model, $x$ and $t$ is the input and output vector, $loss$ is the loss function, $E_{N}$ is the goal sensitive target set from $E$ that will be injected with faults, and $N$ is an empiric upper bound for the number of the faults that the attacker procedure prefers to implement in practice, which is normally much less than the total number of $E$ due to the performance consideration in large CNN model attacks. 

\subsection{Problem Solution}
We employ a fast gradient method similar with FGSM~\cite{goodfellow2014explaining} to solve the optimization problem in Equation (\ref{equ:Problem}), which defines the calculation of the gradient of $loss$ with respect to the element $e_{i}$ as the Equation (\ref{equ:gradients}).% 

\begin{equation}
\begin{aligned}
\label{equ:gradients}
& {\bigtriangledown}_{{e}_{i}}Loss=\left[ \frac{{\delta}{Loss}}{{\delta}{b}_{1}}, \frac{{\delta}{Loss}}{{\delta}{b}_{2}},\dots, \frac{{\delta}{Loss}}{{\delta}{b}_{w}} \right] \\
& where \ \frac{{\delta}{Loss}}{{\delta}{b}_{j}} = loss(f(x;\widetilde{{b}_{j}}),t) - loss(f(x;{b}_{j}),t), \\
& \qquad j = 1, ..., w, \text{and $w$ is the word width}
\end{aligned}
\end{equation}

In the gradient calculation for $e_{i}$, ${\delta}{Loss}/{\delta}{b}_{j}$ $(j = 1, ..., w)$ is calculated as the difference between the loss value of $b_{j}$ and that of $\widetilde{{b}_{j}}$ after it is flipped, which can be used to evaluate the sensitivity of bit $b_{j}$. Generally speaking, attackers could have at least three options to fault the sensitive targets, i.e., the bit-wise injection, word part-wise (part of word), and element-wise (whole word) according to different injection precision. However, it is very hard to precisely fault a bit for a given element by using the voltage and frequency-related fault injections in practice, therefore, the aim of the sensitive target search algorithm just sticks to find the sensitive elements or element parts rather than the sensitive bits. We assume all the bits in an element have the same probability to be faulted, then the mean value of the bit-wise loss gradients $S_{i}$ is used to evaluate the sensitivity of the element $e_{i}$ with $w$ bits ($n$=$w$), or its element part i.e. the mantissa or exponent in this study with $n$ bits as Equation (\ref{equ:perturbedbits}) defined. 

\begin{equation}
\begin{aligned}
\label{equ:perturbedbits}
& S_{i} = \frac{{\sum}_{j=1}^{n} \frac{{\delta}{Loss}}{{\delta}{b}_{j}}}{n} \\
\end{aligned}
\end{equation}

Based on the definition, Algorithm \ref{alg:sensitivityEval} can be employed to evaluate the sensitivity for each target, i.e., the feature map element or element part over all the layers. After all the targets are evaluated with their sensitivities, the top-$N$ of them are selected to form the sensitive target set $E_{N}$ for subsequent fault injections. It should be noted that the values of the gradients at the bits might be negative or positive, and the sum of them means the total expected sensitivity of the target, i.e., the target with negative sensitivities will be discarded in the selection. 

\begin{algorithm}
\renewcommand{\algorithmicrequire}{ \textbf{Input:}}
\renewcommand{\algorithmicensure}{ \textbf{Output:}}
\caption{Sensitivity evaluation algorithm for each candidate target} 
\label{alg:sensitivityEval} 
\begin{algorithmic}[1]
\Require
\textbf{$f$}: Victim CNN model,
\textbf{$x$}: Input vector of CNN,
\textbf{$t$}: Expected output with respect to $x$,
\Ensure
\textbf{$S$}: Sensitivity value array of each element 
\Procedure{Evaluate\_Sensitivity}{$f$,$x$,$t$}
 \State //Initialize \ the \ element \ set
  \State $E \leftarrow$ $allElementsOfAllLayers$($f$) 
  \For{each element $e_{i}$ in $E$}
        \State $ S[i] \leftarrow$ $0$
        \For{each bit $b_{j}$ of $e_{i}$}
            \State $ S[i] \ = \  S[i] + \frac{{\delta}{Loss}}{{\delta}{b}_{j}}$
      
        \EndFor
        \State $ S[i] \leftarrow$ $S[i]/n$ \ \ \ \  // Mean sensitivity for $n$ bits
    \EndFor
    \State \Return $S$
\EndProcedure
\end{algorithmic}
\end{algorithm}

Algorithm \ref{alg:sensitivityEval} shows an example for the sensitivity calculation of the whole element (word-wise), and those for the mantissa and exponent parts are omitted, which are calculated as the mean values of the mantissa and exponent bits sensitivities, respectively. 

\subsection{Sensitive Targets Search Algorithm}
\label{STsalgorithm}
According to the gradient calculation in Equation (\ref{equ:gradients}), the gradient value at a given element $e_{i}$ is related to the values of its bits, $b_{1}$, $b_{2}$, ..., $b_{w}$, which is actually determined by the input feature map of the CNN (e.g., the input image). So, the sensitivity evaluation of the feature map elements as defined in Equation (\ref{equ:perturbedbits}) is also model input-dependent, which can be used in a targeted attack for misleading an original output classification to another specified class, e.g., misleading the recognition for "a panda" to "a monkey". However, an input-independent sensitivity evaluation could be also possible since the target CNN model is trained from a data set composed of all those inputs and such a sensitivity evaluation might be also model structure-correlated. Therefore, this study proposes two algorithms in sensitive target search, one for input-dependent and the other is for input-independent, as Algorithm \ref{alg:SearchAlgorithm} shows.

\begin{algorithm}
\renewcommand{\algorithmicrequire}{ \textbf{Input:}}
\renewcommand{\algorithmicensure}{ \textbf{Output:}}
\caption{Input-dependent and input-independent search algorithms for sensitive targets} 
\label{alg:SearchAlgorithm} 
\begin{algorithmic}[1]
\Require
\textbf{$f$}: Victim CNN model,
\textbf{$x$}: Input vector,
\textbf{$t$}: Expected output with respect to $x$,
\textbf{$N$}: The total element number of the sensitive set
\Ensure
\textbf{$E_{N}$}: Sensitive element set
\Procedure{Input\_Dependent\_Search}{$f$, $x$, $t$, $N$}
  \State $S \leftarrow$ \textsc{Evaluate\_Sensitivity}($f$, $x$, $t$)
  \State $//$ Get top-N elements with significant sensitivities;
  \State $E_{N} \leftarrow$ \textsc{GetTopSet}($f$,$S$,$N$)
  \State \Return $E_{N}$
\EndProcedure
\Procedure{Input\_Independent\_Search}{$f$, $N$}
  \For{each  input   $x$  and  output  $t$}
   \State $S\ = \ S \ + \ $ \textsc{Evaluate\_Sensitivity}($f$, $x$, $t$)
  \EndFor
  \State $E_{N} \leftarrow$ \textsc{GetTopSet}($f$,$S$,$N$)
  \State \Return $E_{N}$
\EndProcedure
\end{algorithmic}
\end{algorithm}

The major difference between the input-dependent and independent algorithms is the way they evaluate the element sensitivities. The input-dependent algorithm selects the sensitive set (top $N$ from the whole set $E$) according to a given input and works for a input-wise attack fashion, while the input-independent algorithm selects the sensitive set according to all the data set being used in training the CNN model and works for a model-wise attack fashion. The former can be more precise for attacking a given input, such as compromising a given victim face recognition, and the latter can be more convenient for attacking a given model to degrade its performance without considering its inputs being processed. 

\subsection{Effect of the Targets Selection}
\begin{figure}[htb]
\begin{center}
\includegraphics[width=1.0\columnwidth]{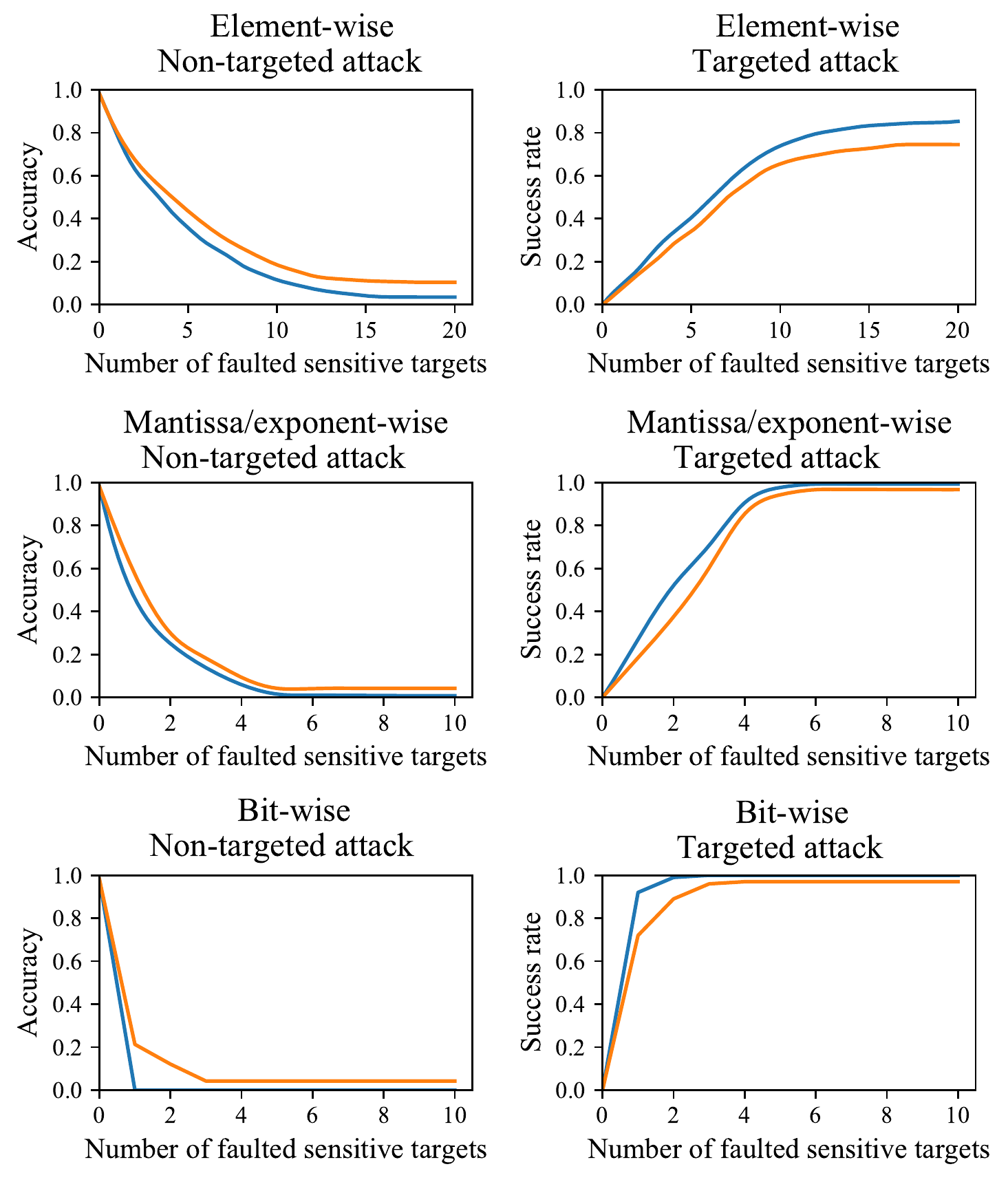}
\caption{The simulation results on Lenet-5 model for evaluating the effects of the targets selection by element-wise, mantissa/exponent-wise, and bit-wise sensitive target set-based attacks. Both the input-dependent (blue line) and input-independent (orange line) sensitive target search algorithms are simulated in targeted attack and non-targeted attack fashion, respectively.}
\label{fig:LossEvo}
\end{center}
\end{figure}

In order to investigate the effect of the sensitive targets search algorithm on the attack efficiency, we simulate the bit-flip faults on the sensitive element set, sensitive element parts set (mantissa and exponent), and sensitive bit set achieved by the method above, respectively, as the example by Lenet-5 model shown in Figure \ref{fig:LossEvo}, to observe the effects on the targeted attack success rate and the non-targeted accuracy degradation. 

The targeted attack means to confuse the CNN to misclassify a class to another targeted class, while the non-targeted attack means just to degrade the model output accuracy without specifying any targeted class for a given input to be misclassified. The simulation randomly flips a bit for each sensitive element or sensitive mantissa/exponent part for $1000$ times, and flips the specified bit for each of the sensitive bit set, after which the mean accuracy or success rate is calculated. As Figure \ref{fig:LossEvo} shows, both input-dependent and input-independent sensitive target search methods (Algorithm \ref{alg:SearchAlgorithm}) are studied for the effect of the targets selection, which actually shows comparable performance with each other in all the three target selection schemes and two attack fashions (targeted and non-targeted). 

For the non-targeted attack fashion, we can see that the accuracy degrades rapidly while the number of faulted sensitive targets ($N$) increases, especially in the bit-wise and mantissa and exponent-wise target cases, which means the bit-wise and mantissa/exponent-wise target selection can gain better performance in accuracy degradation than the element-wise selection. However, the bit-wise and mantissa exponent-wise methods require more precision in fault injection than the element-wise, which makes them harder to realize in attack practice than the element-wise, too. 

For the targeted attack fashion, we can see that the success rate for accomplishing the targeted misclassification also increases with the number of sensitives targets having been faulted, especially in bit-wise and mantissa and exponent-wise target cases. Similar with that in non-targeted attack fashion, the bit-wise and mantissa/exponent-wise target selection can also gain better performance in targeted attack success rate than the element-wise selection, but the latter has better advantages in ease of implementation in faulting practice.

It should be noted that both the input-dependent and input-independent methods show similar performance in all the cases, and as an interesting result to point out, even in targeted-attack, the input-independent attack may also be able to achieve similar performance with the input-dependent in confusing the CNN model running on GPU to misclassify all the inputs to a specific target class, which may be more convenient and efficient to use than the input-dependent method in real attacks since the attacker do not need to first figure out what the current input is.

\section{Fault Injection Parameter Search}
\label{sec:ParamterFind}
%As mentioned above, this study constrains hardware faults to make only one bit-flip on selected sensitive targets, which means a precise fault injection parameter scheme as defined in Table \ref{table:ParameterMeaning} should be addressed for each injection in terms of timing and position (bit) precision. Given sensitive targets logically provided by the method in Section \ref{SearchAlgorithm}, when and where to physically realize a hardware fault injection directly determines the usability and performance of the attack. In this study, a genetic algorithm is proposed to automatically search fault injection parameters available for the attack, which can meet both the timing and position (bit) requirements.
According to the simulation results in Section \ref{SearchAlgorithm}, bit-wise or mantissa/exponent-wise sensitive target faulting may be the most effective. However, implementing such fine-grained attacks is far more difficult than implementing the element-wise due to the limitation of the faulting precision in practice. Consequently, this study employs the element-wise sensitive targets based on which a genetic algorithm is proposed to automatically search fault injection parameters available to implement the attack.

\subsection{Parameters for Timing and Position}
\label{PreAna}
The actual timing points for the attacker to inject faults highly depend on how the CNN models are implemented and scheduled on GPUs. Nvidia GPUs provide the Compute Unified Device Architecture (CUDA) to compile and profile the CNN models, and the GPU running (execution) time until any selected sensitive element of a feature map starts can be tested and calculated by performing the victim model on the GPU with a fixed frequency and voltage, which is the rough value for the parameter $T_{W}$ in Table~\ref{table:ParameterMeaning}.

\begin{figure}[htb]
\begin{center}
\includegraphics[width=0.65\columnwidth]{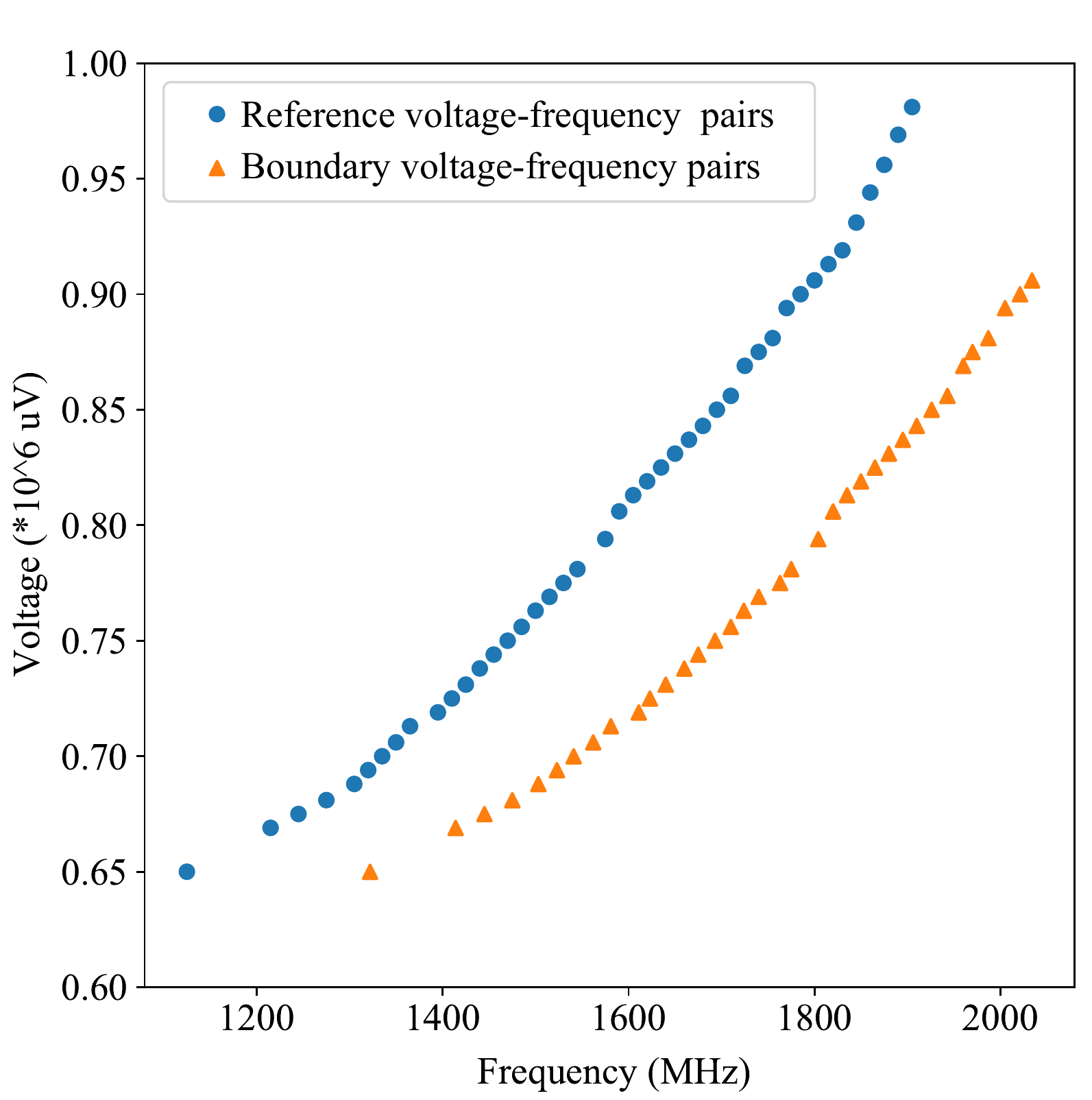}
\caption{The safe voltage-frequency boundary for Nvidia GeForce 1650 \victimdevice. The blue dots indicate the normal voltage-frequency pairs, and the orange triangles indicate the boundary voltages (lowest safe) for the corresponding frequencies vertically projected on x-axis.}
\label{fig:testgpu}
\end{center}
\end{figure}

Compared to the timing, the position precision mainly means the bit position and the number of target bits to be faulted within an element, which is set to $1$ bit in this study as mentioned earlier. However, it is very hard to control the fault to happen at a specific bit position by the voltage and frequency hardware faulting in practice, therefore, the major goal of optimizing the position precision is just to guarantee the faults to happen as $1$-bit flips within specific sensitive target, in this study, in which the faulting voltage and frequency pair ($V_{l}$ -$F_{h}$) should be firstly investigated. As Figure \ref{fig:testgpu} shows, an Nvidia GeForce 1650 GPU prefers reference voltage and frequency pairs as the blue dots indicate, based on which we can test and acquire the boundary voltages (orange triangles) that are the lowest values for GPU to run under each paired reference frequency, respectively. So, the voltage-frequency pairs between the reference pairs and the boundary voltages are safe, but those below the boundary voltages are ether overclocking (horizontal direction) or undervoltaging (vertical direction) since the voltages are insufficient for GPU to run under the paired frequency, in which the timing violation-based hardware faults may be probably injected. 

\begin{figure}[htb]
\begin{center}
\subfigure[The number of faulted bits with different $V_{l}$-$T_{d}$ pairs (the frequency offset is $235$MHz in this figure).]{\label{fig:testduration1}
\includegraphics[width=0.8\columnwidth]{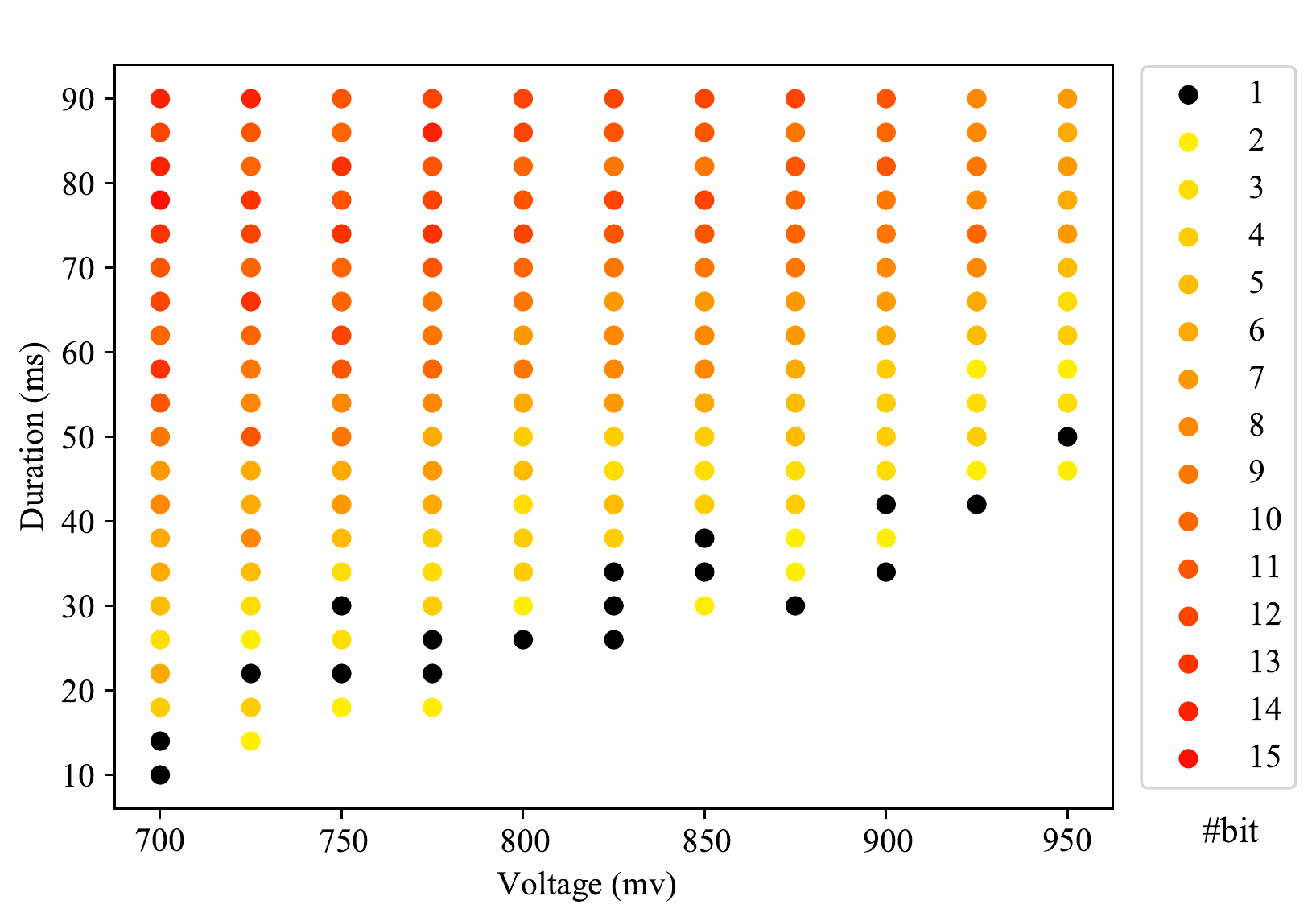}}
\subfigure[The number of faulted bits with different $f_{h}$-$T_{d}$ pairs (the voltage is $710$mV in this figure).]{\label{fig:testduration2}
\includegraphics[width=0.8\columnwidth]{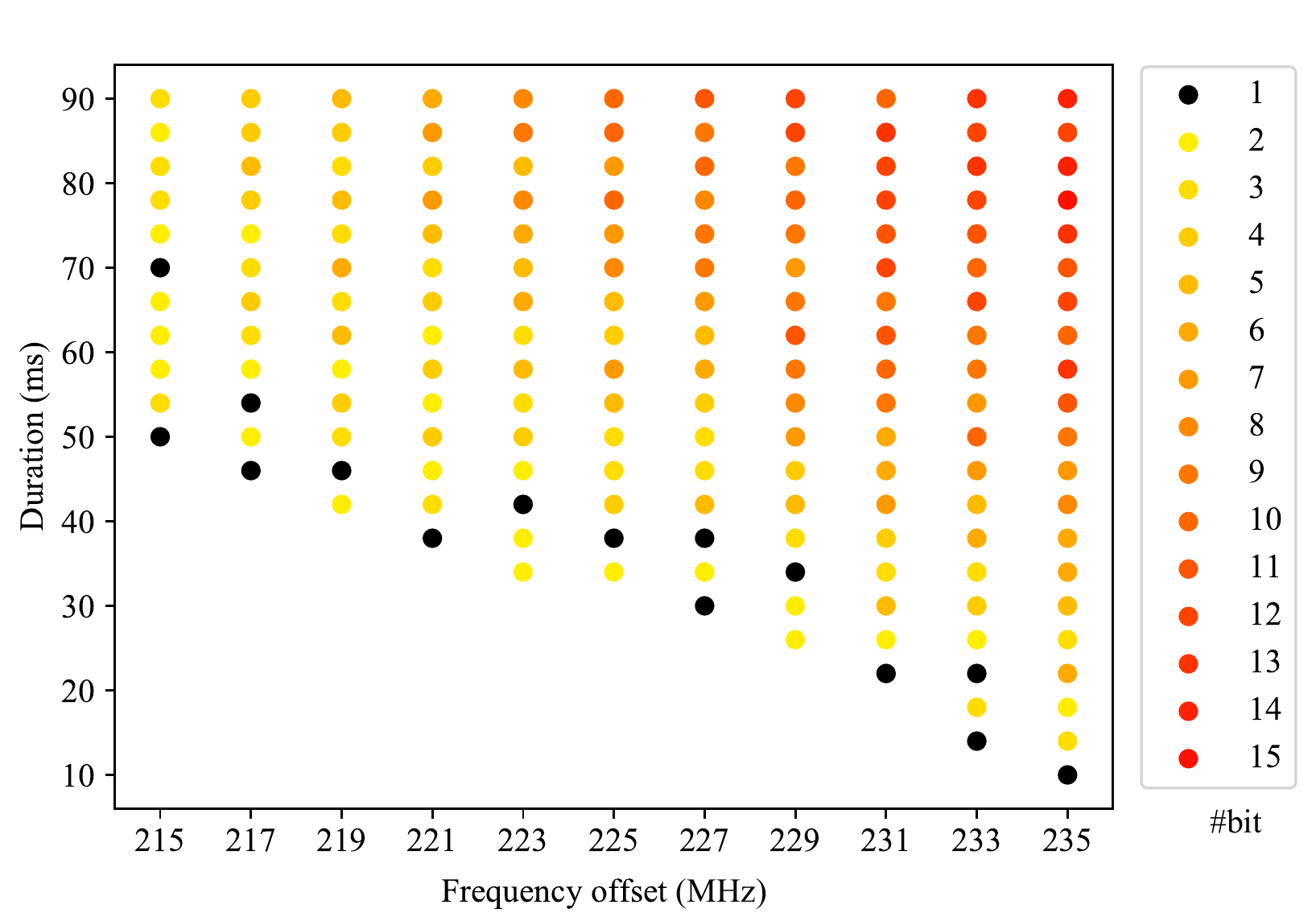}}
\caption{The average number of faulted bits with various fault injection parameters, from which we can see that $T_{d}$ has a remarkable influence on the number of faulted bits.}
\label{fig:testduration}
\end{center}
\end{figure}

Another parameter critical for controlling the faulting bits is the period $T_{d}$ during which the faulting voltage-frequency pair holds. Figure \ref{fig:testduration} illustrates the experimental results about the number of faulted bits with various injection parameters, in which we test $5$ times for each parameter and plot the average number of faulted bits into this figure. As shown in Figure \ref{fig:testduration}, the parameters that are easy to implement the $1$-bit fault injection are close with each other in values and the number of faulted bits is highly influenced by $T_{d}$, which shows a good locality characteristic and is good for a possible search algorithm-based optimal attack parameter determination.

\subsection{Comparing Fault Injection Strategies}
Generally, there are three types of fault injection strategies since both the voltage and frequency can be manipulated, i.e., 
\begin{itemize}
    \item Only undervoltaging.
    \item Only overclocking.
    \item Simultaneous manipulation with low voltage and high frequency.
\end{itemize}

We test the three strategies on Nvidia GeForce 1650 GPU by trying each $V_{l}$-$F_{h}$ pair for $1,000$ times and conclude the rate of different behaviors of GPU. We observe that both the undervoltaging-only and overclocking-only strategies can cause the statuses of a crash or no response in a very high ratio, and the simultaneous manipulation with voltage and frequency can result in a high success rate of fault injections, which is preferred in this study for high efficiency. Among all the fault injection $V_{l}$-$F_{h}$ pairs, we find the highest successful injection rate occurs when the voltage is $710$mV and the frequency offset is $235$MHz.

\subsection{Refining Fault Injection Parameters}
\begin{figure}
\begin{center}
\includegraphics[width=0.9\columnwidth]{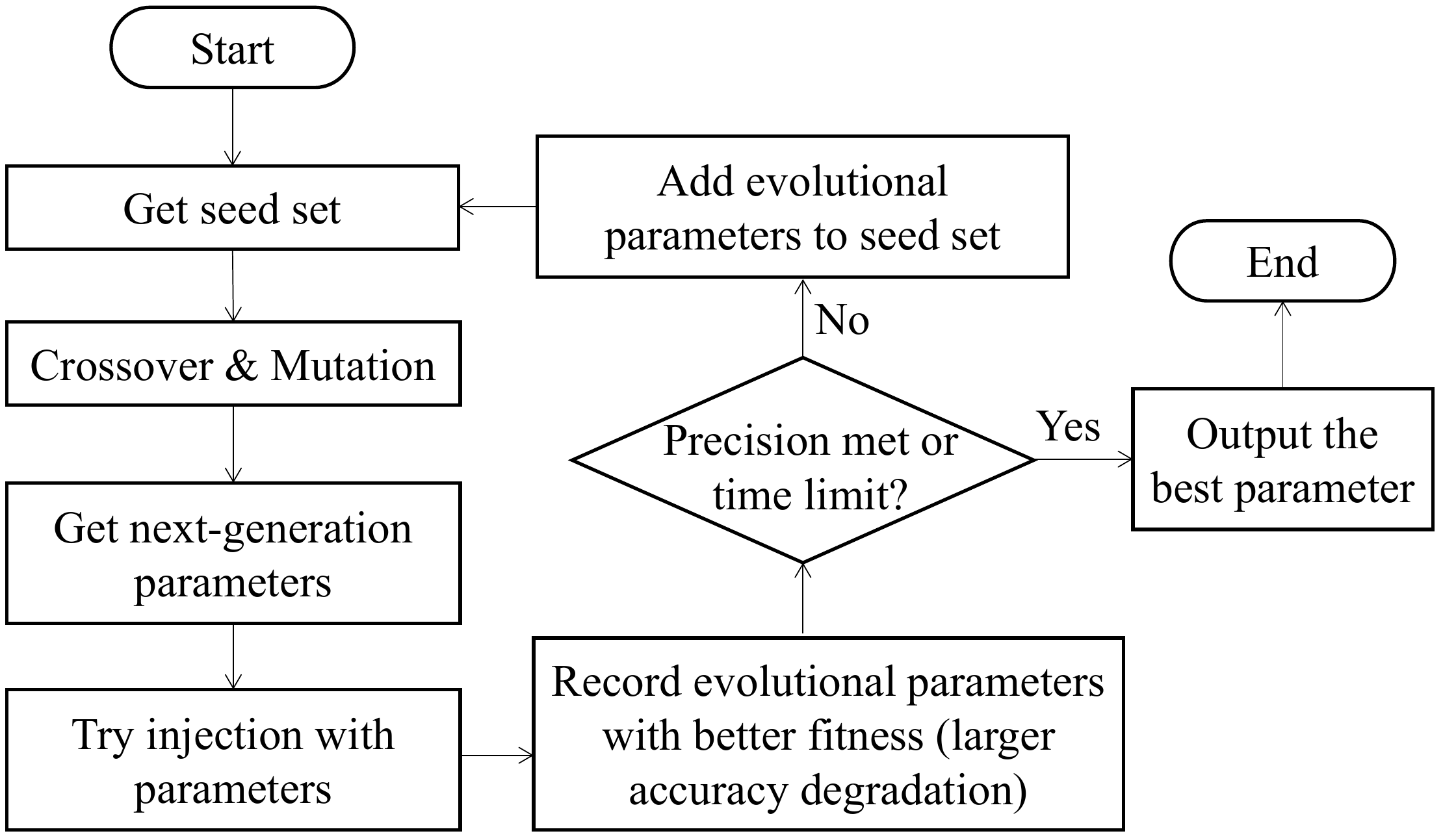}
\caption{The genetic algorithm for refining the fault injection parameters.}
\label{fig:refining}
\end{center}
\end{figure}

Because of the runtime variation and parallel structure of GPU, there is slight non-determinism in the timing of computation of the certain GPU unit. In order to meet the timing and position requirements and improve the fault injection precision, we employ a genetic algorithm to refine the parameter for a better injection precision, as shown in Figure \ref{fig:refining}.

At the beginning of the algorithm, an initial seed set of fault injection parameters is selected based on the rough parameters determined as mentioned above, following which crossover and mutation processes will be executed on the initial seed set to create new parameters for the subsequent fault injection trials. Each seed is encoded as a sequence of $(F_{h},V_{l},T_{w},T_{d})$. 
%, in which $F_{h}$ and $V_{l}$ are the glitch frequency and voltage respectively, and $T_{w}$ and $T_{d}$ are the timing to wait and the duration of glitch for the fault injection respectively. 
The crossover operation randomly exchanges a sub sequence of two sequences (seeds), and in the mutation process, the parameter $F_{h}$, $V_{l}$, $T_{w}$, and $T_{d}$ is increased or decreased by step of $1MHz$, $10mv$, $1ms$, and $1ms$ respectively by the probability of $0.5\%$. We monitor the accuracy degradation as the fitness of the new parameters. 
%Those parameters with larger accuracy degradation will be added to the seed set. During this process, 
In this algorithm, the tournament selection is utilized as the selection method for the next generation. However, this may drop in the local optimum. In our implementation, the parameter with less accuracy degradation will be selected if a parameter that should be selected in this time has already been selected into the seed set more than $20$ times. After the new seeds are added to the seed set, the genetic algorithm flow will cycle again until the injection parameters reach the expected precision or the cycle reaches the time limit. This algorithm enables us to automatically search for the effective fault injection parameters for the given sensitive targets provided by the algorithm presented in Section \ref{SearchAlgorithm}.

\section{Experimental Evaluation}
\label{Experiments}
We verify \papertitleabbr\ attack in attacking three Nvidia GPUs for four widely used CNNs with MNIST, CIFAR-10, and Yale face data sets. %The experiments are conducted on our local machines, however, as mentioned in the threat model in Section \ref{sub:threatmodel}, the attacks would be a potential risk for the GPU cloud whose guest VMs are protected as the hypervisor have the power to manipulate the voltage and frequency of target GPUs.

\subsection{Experiment Setup}
\begin{figure}
\begin{center}
\includegraphics[width=0.9\columnwidth]{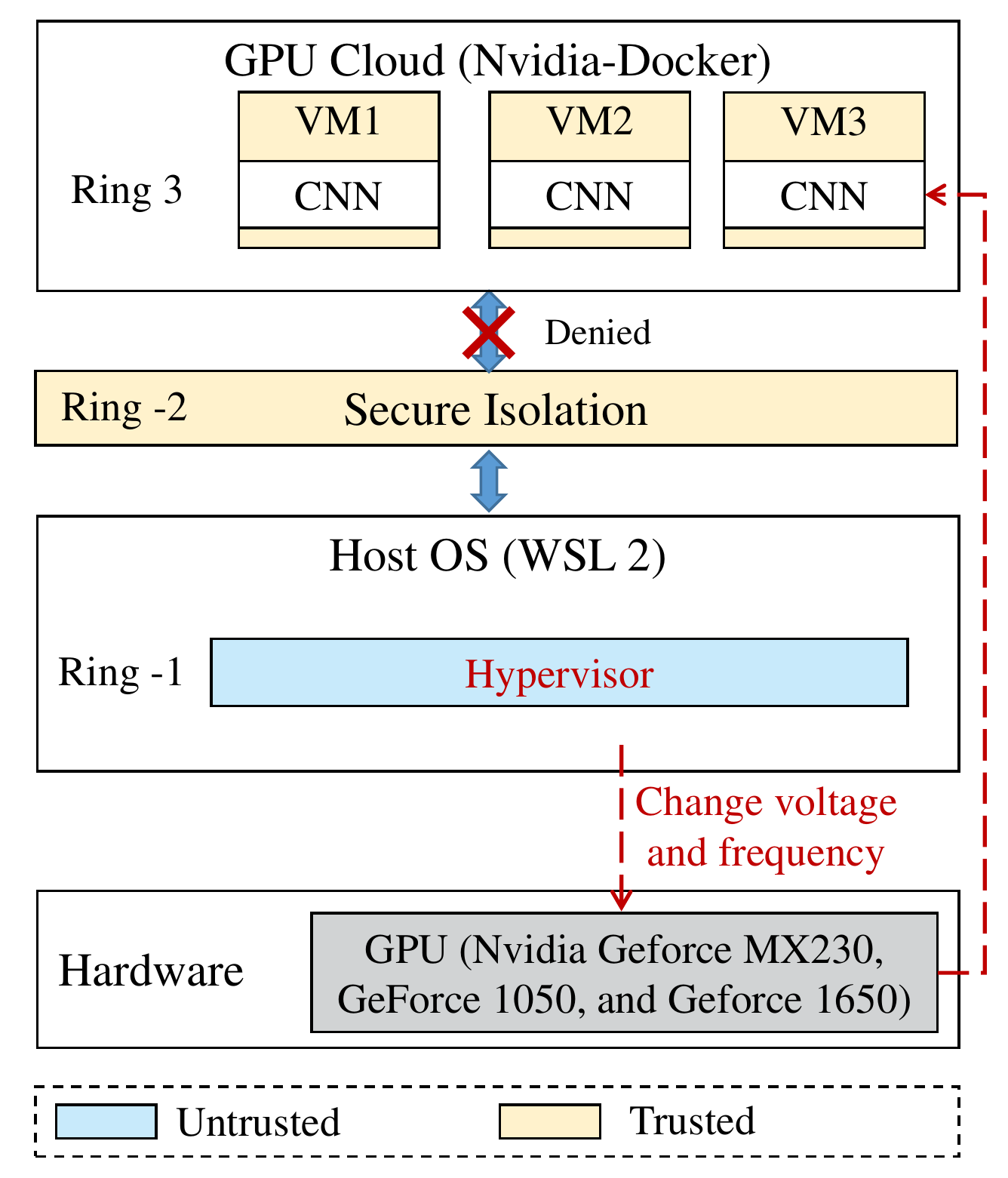}
\caption{The experiment setup.}
\label{fig:ExperimentSetup}
\end{center}
\end{figure}

Figure \ref{fig:ExperimentSetup} illustrates the experiment setup in this study. There are no open-source projects available for secure isolation designs, therefore we refer to the CloudVisor architecture~\cite{zhang2011cloudvisor} to build our experiment environment. The main idea of CloudVisor is to introduce a tiny security monitor underneath the Virtual Machine Monitor (VMM) using nested virtualization and provides protection to the hosted VMs~\cite{zhang2011cloudvisor}. In our experiments, a trusted layer are created between the hypervisor and VMs, which intercepts and encrypts all the communications between the hypervisor and VMs to guarantee VMs are securely isolated from hypervisor. %The integrity of the trusted layer is guaranteed by the Trusted Platform Module (TPM) through authenticated booting procedure.

In the traditional VM environments like VMware and VirtualBox, we can not assign the CNN to run on the GPUs as they do not virtualize the GPU. In this work, we utilize the Windows Subsystem for Linux (WSL 2) to build our experiment environments. The WSL 2 enables us to run a Linux (Ubuntu) environment directly on Windows OS and also have supports to GPU-accelerated applications. In order to create a cloud environment, we utilize the Nvidia-docker to create multiple containers for CNNs on the Ubuntu system, and each container is a guest VM. %This experiment setup is an example, 
%for demonstration, as some advantage security properties of CloudVisor are not implemented, 
%but we believe that the attack would be efficient on any GPU cloud as hypervisors have a full control over the cloud devices.
%Compared to the WSL 1 that  coverts Linux kernel APIs to the equivalent Windows kernel APIs, the Linux system in WSL 2 is run as a “Lightweight Utility VM” under Hyper-V virtual machine technology and the entire file system is implemented in a single vhdx file. This prevents the administrator (elevated) Windows processes from modifying a sensitive Linux files legally. Of course, the elevated Windows processes can actually alter the sensitive Linux files in the current situation by the following three steps: 1) covert the file system of WSL 2 to WSL 1 file system; 2) alter the sensitive Linux files; 3) overt the polluted files to the WSL 2 file system. 

\emph{Data set.}
The MNIST is a handwritten digits data set from $0$ to $9$, which has a training set of $60,000$ examples and a test set of $10,000$ examples. The CIFAR-10 data set contains $60,000$ RGB images in size of $32 \times 32$, in which $50,000$ images are training set and $10,000$ images are test set. The images are drawn evenly from $10$ classes. The Yale face is a widely used data set for face recognition, which contains $165$ grayscale images in GIF format of 15 individuals.

\emph{CNN models.}
\label{ModelDetail}
Four widely-used CNN models are chosen as the attack targets, including Lenet-5, AlexNet, ResNet-18, and MobileNet. %The weight parameters are full-precision and implemented on Nvidia GPUs.

\emph{Hardware and software configurations.}
Three widely-used Nvidia GPUs are chosen as the attack targets, including GeForce MX230, GeForce 1050, and GeForce 1650. The configurations of experiment devices are shown in Table \ref{table:hardcon}. All of the three devices are equipped with the Windows OS and the kernel version is 10.0.20262.1010. The kernel version of the Ubuntu VM in WSL 2 is 4.19.128.

\begin{table*}[htbp]
\caption{The device configuration.}
\centering
\small
\begin{tabular}{l|lll}
\hline
GPU & Nvidia GeForce GTX 1050 & Nvidia GeForce GTX 1650 & Nvidia GeForce MX230 \\
\hline
CPU & Intel Core i5-8300H & Intel Core i5-9400F & Intel Core i5-1035G1 \\
Motherboard & CFL Freed\_CFS & ASUS PRIME B365M-K & DELL 0HP5Y9 \\
PSU & LG PABAS0241231 - 41167 & Huntkey HK500-16FP G1 & BYD DELL 1VX1H02 - 983 \\
\hline
\end{tabular}
\label{table:hardcon}
\end{table*}

\subsection{Non-Targeted Attack}
The non-targeted attack aims to degrade the original model output accuracy without specifying the target misclassified class for given inputs. To verify the effects of the \papertitleabbr\ attack on reducing the model output accuracy, we first train the Lenet-5, AlexNet, and ResNet-18 models using MNIST data set, train a ResNet-18 model using CIFAR-10 data set, and train a Mobilenet model using Yale face dataset. The baseline accuracy of these models is measured, which can be found in Table \ref{table:mnistresults}. All of the models have a relatively high accuracy (more than $97\%$). It should be noted that the fault injection number $N$ cannot be very large because the faults need to be induced during a very fast inference procedure. According to the inference time and duration of the fault injection glitches, $N$ is set to an empiric number of 10 in this study.

\begin{table}[htbp]
\caption{The degraded model accuracy under different attack methods. Accuracy in the last column is for the degraded results after input-dependent (Depend.) and input-independent (Independ.) attacks, respectively.}
\centering
\small
\begin{tabular}{p{1.6cm}|p{0.7cm}|l|p{1.1cm}p{1.1cm}}
\hline
\multirow{1}{*}{Model} & \multirow{1}{*}{Base } & \multicolumn{1}{c|}{\multirow{2}{*}{GPU}} & \multirow{1}{*}{Accuracy} & \multirow{1}{*}{Accuracy}\\
(Data Set) & Accur. &  & (Depend.) & (Independ.)\\
\hline
\multirow{1}{*}{Lenet-5} & \multirow{3}{*}{98.2\%} & MX230 & 30.4\% & 38.6\%\\
(MNIST) & & GeForce 1050 & 22.2\% & 27.7\%\\
 & & GeForce 1650 & 19.9\% & 26.3\%\\
 \hline
 \multirow{1}{*}{AlexNet} & \multirow{3}{*}{98.7\%} & MX230 & 36.2\% & 42.7\%\\
 (MNIST) & & GeForce 1050 & 27.8\% & 33.4\%\\
 & & GeForce 1650 & 26.3\% & 34.0\%\\
 \hline
 \multirow{1}{*}{ResNet-18} & \multirow{3}{*}{99.1\%} & MX230 & 40.2\% & 45.1\%\\
 (MNIST) & & GeForce 1050 & 31.2\% & 37.6\%\\
 & & GeForce 1650 & 28.9\% & 34.7\%\\
 \hline
 \multirow{1}{*}{ResNet-18} & \multirow{3}{*}{97.6\%} & MX230 & 43.1\% & 48.3\%\\
 (CIFAR-10)  & & GeForce 1050 & 29.5\% & 36.2\%\\
 & & GeForce 1650 & 31.1\% & 36.0\%\\
 \hline
 \multirow{1}{*}{Mobilenet} & \multirow{3}{*}{98.8\%} & MX230 & 36.4\% & 46.1\%\\
 (Yale)  & & GeForce 1050 & 27.3\% & 35.2\%\\
 & & GeForce 1650 & 26.1\% & 35.8\%\\
 \hline
\multicolumn{3}{c|}{Average} & 30.8\% & 37.2\%\\
 \hline
\end{tabular}
\label{table:mnistresults}
\end{table}

\subsubsection{Input-Dependent Attack}
We first evaluate the input-dependent attacks on the four victim models and measure the average output accuracy, as demonstrated in Table \ref{table:mnistresults}. On MNIST data set, we can see that the \papertitleabbr\ attack can reduce the accuracy of LeNet-5 model from $98.2\%$ to $30.4\%$ (MX230), $22.2\%$ (GeForce 1050), and $19.9\%$ (GeForce 1650) respectively, reduce the accuracy of AlexNet model from $98.7\%$ to $36.2\%$ (MX230), $27.8\%$ (GeForce 1050), and $26.3\%$ (GeForce 1650) respectively, and reduce the accuracy of ResNet-18 model from $99.1\%$ to $40.2\%$ (MX230), $31.2\%$ (GeForce 1050), and $28.9\%$ (GeForce 1650) respectively. On CIFAR-10 data set, the average accuracy degradation of ResNet-18 model is $63.0\%$ over all the three GPUs. On Yale face dataset, the average accuracy degradation of Mobilenet model is $68.9\%$ over all the three GPUs.
 The average degradation is $67.7\%$ over all the four models, the three data sets, and the three GPUs. The reason why the performance on MX230 GPU is lower than the other two is mainly because the fault injection parameter space of MX230 available for the voltage and frequency to be selected from is very limited. 

\subsubsection{Input-Independent Attack}
Compared to the input-dependent attack, the input-independent attack does not depend on the given input to determine where and when to inject faults, which is determined in advance via testing the data set before attacking. Table \ref{table:mnistresults} illustrates the results of the model output accuracy with respect to the input-independent attack. On MNIST data set, we can see that the \papertitleabbr\ attack can reduce the accuracy of LeNet-5 model from $98.2\%$ to $38.6\%$ (MX230), $27.7\%$ (GeForce 1050), and $26.3\%$ (GeForce 1650) respectively, and reduce the accuracy of AlexNet model from $98.7\%$ to $42.7\%$ (MX230), $33.4\%$ (GeForce 1050), and $34.0\%$ (GeForce 1650) respectively, and reduce the accuracy of ResNet-18 model from $99.1\%$ to $45.1\%$ (MX230), $37.6\%$ (GeForce 1050), and $34.\%$  (GeForce 1650) respectively. The average degradation is $62.5\%$ over all the three models and three GPUs. On CIFAR-10 data set, the average accuracy degradation of ResNet-18 model is $57.4\%$ over all the three GPUs. On Yale face dataset, the average accuracy degradation of Mobilenet model is $59.8\%$ over all the three GPUs. The average degradation is $61.3\%$ over all the four models, the three data sets, and the three GPUs, which is a little lower than that of the input-dependent method ($67.7\%$), but the input-independent attack may be easier to use in practice.

\subsection{Targeted Attack}
\begin{figure}[htb]
\begin{center}
\includegraphics[width=0.9\columnwidth]{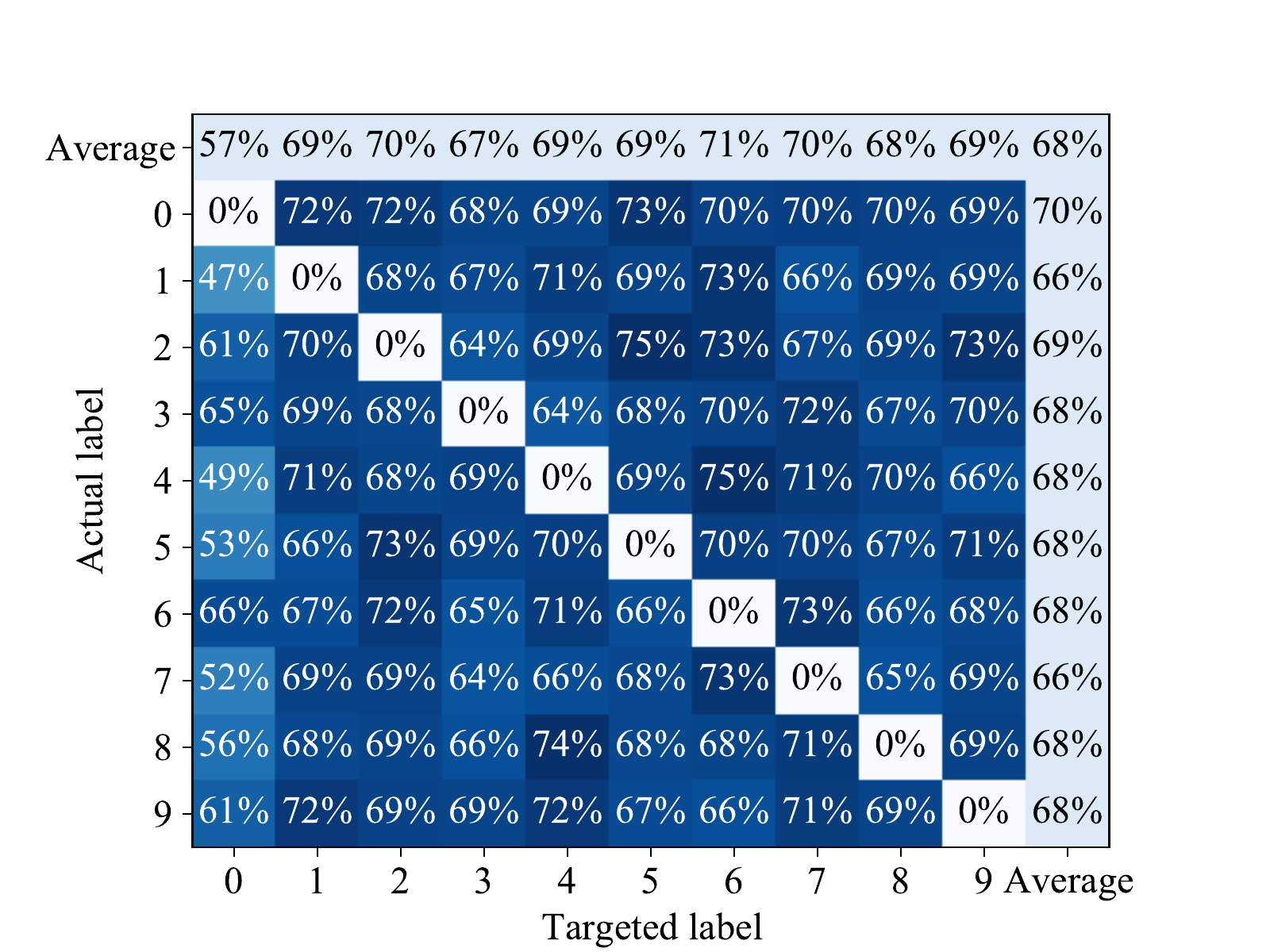}
\caption{The confusion matrix of targeted attack (input-dependent) on Lenet-5 model
with MNIST data set.}
\label{fig:testtarget}
\end{center}
\end{figure}

The targeted attack aims to confuse the CNN to misclassify a label (class) to another targeted label (class). %, i.e., for a given input, it can be misclassified as another label (class) different from the actual one. 
The confuse matrix for the Lenet-5 model on the GeForce 1650 GPU under input-dependent attack is shown in Figure \ref{fig:testtarget} as an example, in which each of the $10$ labels (rows) is evaluated by the success rate to be misclassified as another targeted label (columns) in $100$ times of attacks. The diagonal elements mean the correct classifications the original model should output, which are not hit throughout the attacks. From the results, we can see that the worst cases mainly occurs when misclassifiying numbers to the target label of $0$, whose average success rate is about $56.7\%$, while the others have a mean success rate of $69.2\%$. Such a performance variation is similar with what the adversarial example attacks imply~\cite{kwon2018random}, the reason of which may be because the digit $0$ is intrinsically different from others in handwriting and hard to confuse. The entire average success rate of the targeted attack of \papertitleabbr\ attack is $67.9\%$ over all the cases, which shows about $30.1\%$ lower than that of the state-of-the-art adversarial example attacks (about $98\%$). It is mainly because the adversarial example attacks can precisely revise every bit of the input image and iterate many times for a better result but the fault injection attacks cannot have a comparable injection precision and can hardly iterate within the period of the target inference process.%, unless it is inferred again by the user. 

\subsection{Summary of Experiments}
The experiments verify that the \papertitleabbr\ attack successfully exploits the disclosed vulnerability of this study to make CNN models untrusted on multi-tenant shared Nvidia GPUs, in both non-targeted attack and targeted attack fashions. For non-targeted attack, the input-independent method can also achieve similar accuracy degradation with the input-dependent, which may be easier to use in practice since it does not need to figure out what the current input is. For a targeted attack, the best result based on the input-dependent method shows that the \papertitleabbr\ attack can also achieve acceptable performance in comparison with the state-of-the-art adversarial example attacks. Compared to the simulation results, the actual performance after using the real hardware faults looks lower, the reason of which is mainly due to the precision errors in conducting the voltage and frequency fault injections in term of timing and position.%, which may be further investigated to improve based on the current method illustrated in Section \ref{sec:ParamterFind}.

\section{Discussion}
\label{Dissicution}

\subsection{Potential Impact}
To the best of our knowledge, \papertitleabbr\ is the first to strike through the secure isolation of GPU clouds and attack DNN via GPU hardware faults. It does not rely on any access violations enforced by the current secure designs. Besides, such attack fashions are free of input adversarial samples and could be deployed more stealthily and usable in an input-independent manner. 

In addition to DNNs, based on the previous studies using DVFS vulnerabilities to attack SGX and TrustZone~cite{xxx},
some encryption schemes such as AES and RSA accelerated by GPU clouds~\cite{nishikawa2017implementation, razaque2018integration} could be also vulnerable to the \papertitleabbr\ attack.
%Field Programmable Gate Array (FPGA)-based accelerators~\cite{gankidi2016fpga}, Google Tensor Flow Processor (TPU)~\cite{gordienko2020scaling}, and Intel Movidius Myriad X~\cite{ionica2015movidius, tsimpourlas2018design}, et al., are also playing important roles in accelerating DNNs. For preserving power efficiency, similar voltage and frequency scaling technologies with those widely used in CPUs and GPUs may also be employed in these architectures, which may make them vulnerable to \papertitleabbr\ attack.

%Besides, according to our communications with the security team of Nvidia, they are actually implementing the GPU-TEE. In GPU-TEE, the root user is considered as an untrusted party, which means that the adversaries are rejected to manipulate the DNNs protected by the GPU-TEE even they have the root privilege. However, \papertitleabbr\ attack would be workable on GPU-TEE as long as the voltage and frequency of the victim GPUs can be operated with software.

%It should be also noted that the \papertitleabbr\ attack could be very dangerous in cloud services since more and more DNNs are provided through cloud-based services shared between different users~\cite{armbrust2010view}. In this scenario, multiple users may share the same hardware resources, and the \papertitleabbr\ attack from one user may directly affect the DNN services of another user.

\subsection{Limitations}
\label{sec:Limitation}
First of all, the success rate of the targeted attack or the accuracy degradation of the non-targeted attack of the \papertitleabbr\ attack may be lower than some software attacks, such as adversarial attacks because of the limited precision and possible failures or variations of the hardware faults. 
%Secondly, this attack requires the privilege to manipulate the voltage and frequency via software drivers provided by the GPU vendors, which may be normally in a system rather than a user privilege. The requirement suggests that the adoption of the attack is subject to the attacker capacity of the system control, which means the untrusted third-party hardware platforms such as cloud services or those having limited resources defending against Advanced Persistent Threat (APT) attackers that have enough system controls may be the most important sources of \papertitleabbr\ attacks. Nevertheless, the proposed vulnerability and  attack in this study is still very important to call for the necessary attention of the academia and industry in a hardware-involved whole-system security consideration, as an innovative addition to the current software or model defect-based attacks for DNNs.

\subsection{Potential Countermeasures}
The potential countermeasures against the proposed security threat for GPU clouds could be realized in the following aspect. 

\textbf{Restricting DVFS}.
The first potential measure may be just to restrict DVFS manipulation from free adjustment of its voltage and frequency when disabling DVFS is not a valid option. Binding the working voltages with fault-tolerant paired frequencies in a registered table and enforcing such a voltage-frequency pair each time DVFS is required to tune the status by the operating systems. However, the countermeasure should be implemented by the hardware logic on GPUs otherwise the hypervisor may still be able to compromise it if it is realized on drivers, which imposes high cost but still to be theoretically guaranteed there will be no hardware faults in all kinds of working conditions.

\textbf{Secure isolation of driver}.
Compared to hardware revisions, the software secure isolation designs with trusted device driver control included are relatively low-cost and easy to scale to other device controls, such as memory for avoiding rowhammer~\cite{seaborn2015exploiting}, CPU for avoiding VoltJockey~\cite{qiu2019voltjockey}, and other potential future hardware faults-based attacks. 

Therefore, different from most of the state-of-the-art secure isolation designs, such a new design may enforce strong security policies to restrict device drivers from influencing guest VM. Proposing an example for the idea of the design, 
a secure isolation extended upon 
Graviton~\cite{volos2018graviton} could be considered. Unlike method in Graviton that protect specific sensitive resource from untrusted driver, we suggest that the system should force all guest and hypervisor to verify identity before manipulating device using driver. This can be achieved by trusted execution environments(TEE) owned by Guest VM.

\begin{figure}[htb]
\begin{center}
\includegraphics[width=0.9\columnwidth]{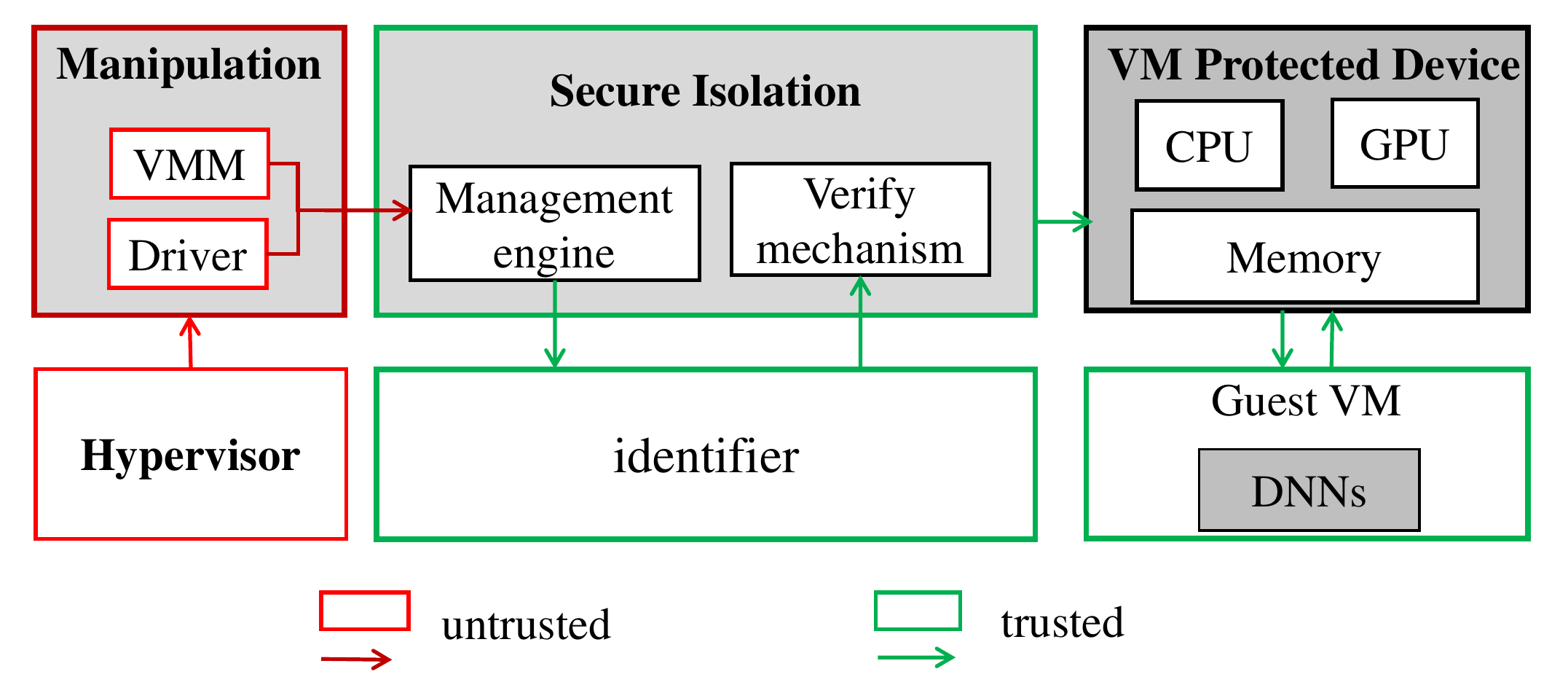}
\caption{Secure architecture of GPU clouds with untrusted hardware. We add an identifier on ring -3 and it can only be accessed by platform management engine and verified through hardware security method such as PUF. it will deny all request from untrusted component including hypervisor, VMM, and driver.}
\label{fig:securemodel}
\end{center}
\end{figure}

\textbf{Secure architecture of GPU clouds with untrusted hardware}.
However, there are some work(~\cite{qiu2019voltjockey},~\cite{qiu2019voltjockeySGX}) that prove that TEE can also be breakthrough and remain untrusted. But we still need a identifier or key to build a strong secure isolation and verify the guest.
So it leads to a new problem that how to generate a identifier with no trust of hypervisor, TEE and hardware. We suggest a potential solution can be provided by special store region only visible to platform management engine(ring -3) and the identifier can verify through hardware security method such as Physical Unclonable Function(PUF). 
We believe such method will be a root countermeasure for such threat of hardware faults. Figure \ref{fig:securemodel} shows a secure architecture of GPU clouds with untrusted hardware. it use the identifier generated by related hardware security method. After system hardware resource management, a identifier is generated to verify the guest. The identifier can only be accessed by platform management engine(ring -3) and will be verified through hardware security method mentioned above when anyone wants to manipulate device(including memory access/driver manipulate). The request will only be granted if the identifier verify the guest. This architecture is based on the security of platform management engine and can isolate any threat from untrusted hypervisor and regulation of hardware.
%The monitor can either run on GPUs or on CPUs.

\section{Related Work}
\label{RelatedWork}
%This section briefs the most related work to this study including the DNN attacks, the voltage and frequency related attacks, as well as the current exposed vulnerabilities of GPUs.
This section briefs the most related work to this study including the voltage and frequency related attacks and the current exposed vulnerabilities of GPUs.

\subsection{Hardware Faults-Based Vulnerabilities}
Hardware faults are widely used to attack computing systems especially for encryption key analysis or information theft. Hardware faults can be injected by hardware tools and devices, such as overclocking~\cite{delvaux2014fault}, undervoltaging~\cite{barenghi2009low, barenghi2010low}, overtemperature ~\cite{hutter2013temperature}, laser and ion radiation \cite{courbon2014adjusting, seifert2010radiation}, or electromagnetic~\cite{dehbaoui2012electromagnetic}, by software~\cite{seaborn2015exploiting, gabor2019high, endo2011chip}, or by simulations~\cite{viera2019simulation, eslami2020survey}. Compared to the hardware-based fault injections relying on hardware tools or devices, which have restricted potential in the capacity of network attacks, the software-based fault injection techniques may have great security threats to cyberspace since the attack might be conducted via network under specific conditions. Besides the aforementioned row-hammer vulnerability, in recent years, the hardware vulnerabilities caused by the implementation defects in DVFS technology have drawn remarkable research attentions, which provides the adversaries attack windows via software drivers under given privileges. The root cause of such vulnerabilities is the timing constraint violations of the synchronous sequential circuits induced by inappropriate working frequency and voltage configurations to the hardware.

Tang et al. disclosed a software-controlled frequency fault-based vulnerability for ARM TrustZone by overclocking techniques~\cite{tang2017clkscrew}, which can make the Trusted Execution Environment (TEE) of ARM CPUs untrusted without the help of any software vulnerabilities.

Qiu et al. proposed a software-controlled voltage-based hardware fault injection attacks on commercial ARM processors~\cite{qiu2019voltjockey}. Besides, Qiu et al.~\cite{qiu2019voltjockeySGX}, Murdock et al.~\cite{murdock2020plundervolt}, and kenjar et al.~\cite{kenjar2020v0ltpwn} demonstrated that the software-controlled voltage-based hardware fault injection attacks are also feasible on Intel processors. They manipulated the processor voltage with software interfaces exposed by the DVFS technology instead of using auxiliary hardware units. 

\subsection{GPU Vulnerabilities}
Currently disclosed vulnerabilities related to GPUs can be mainly classified as three categories, i.e., the side channel vulnerabilities, circuit fault-related vulnerabilities, and software vulnerabilities. 

There are cache memories integrated into modern GPU architectures, which is verified to be vulnerable to cache side channel attacks, such as prime-probe~\cite{naghibijouybari2018rendered}. Besides, timing side channel attacks utilize the execution time differences between different instructions to guess secret data, which are also successfully verified available on GPUs to leak critical information~\cite{karimi2018timing, jiang2016complete, luo2018gpu}. Additionally, attackers can also employ power side channel techniques to gather power footprints while the victim procedure is in execution, which can support the power differential analysis to fulfill the attacks to leak critical information from \victimdevice~\cite{luo2015side, luo2018power}. The disadvantage of side channel attacks is that they cannot modify the internal states of GPUs and therefore cannot be employed to compromise the processing results of the executed routines. 

GPUs are also vulnerable to the circuit fault-related attacks. Sabbagh et al. \cite{Majid2020Novel} proposed to attack AES encryption executed on AMD GPU by using the voltage-induced hardware faults, which does not support GPU accelerators for DNN attacks. 

In addition to the vulnerabilities relatively belonging to the hardware design side, GPUs are also vulnerable to the conventional software defects-based attacks, such as heap and stack buffer overflow~\cite{miele2016buffer, erb2017dynamic, lee2014stealing}, and uninitialized memory~\cite{di2016study}. Such attacks mainly rely on software vulnerabilities and are also easy to be fixed on the software side without high-cost  hardware revisions, which is therefore not considered as typical hardware vulnerabilities within the scope of this study.

\section{Conclusion and Future Work}
\label{Conclude}
In this study, we explore a hardware fault by exploiting the Dynamic Voltage and Frequency Scaling (DVFS) technology. It can be exploited to make DNN (CNN for experiments) on shared GPU cloud untrusted non-intrusively. It can threat the state-of-the-art secure execution environment for guest applications by enforcing strong security policies to isolate the untrusted hypervisor from the guest virtual machines. A sensitive targets search algorithm and a genetic fault injection parameter search algorithm are proposed to address the major challenges in the attack target selection and the fault injection precision in terms of timing and position. Experiments on commodity Nvidia GPUs with four widely used CNNs verify the efficiency of the attack. Furthermore, the proposed attack can also support a high-performance targeted attack. This study reveals a potentially severe threat existing in secure isolation of shared GPU cloud and we propose a secure arichitecture with untrusted hardware as a countermeasure%The security risk is due to the design that the dynamic voltage and frequency of GPUs are not appropriately regulated and transient abnormal voltage-frequency glitches may induce hardware circuit timing constraint variations causing incorrect bit flips. A sensitive targets search algorithm and a genetic fault injection parameter search algorithm are proposed to address the major challenges in the attack target selection and the fault injection precision in terms of timing and position. Experiments on commodity Nvidia GPUs with four widely used CNNs verify the efficiency of the attack. Furthermore, the proposed attack can also support a high-performance targeted attack. As an innovative addition to the current DNN attacks from the perceptive of hardware vulnerability, this study reveals a potentially severe threat existing in hardware accelerator designs and also shows its notable impacts to the security of the DNN-related systems on shared GPU clouds.
In future work, we will study impact of hardware faults on the training phase of DNN models on shared GPU cloud. As the bit-flip faults induced by DVFS will also influence the training and can be exploited to replace the legitimate model of the target device with the poisoned model. To specify the sensitive targets of DNN models in training phase is more hard to implement and we need further study on the exploitation.

%In this study, we disclose a voltage and frequency fault-related risk verified on Nvidia GPUs that can be exploited to make DNN (CNN for experiments) on shared GPU cloud untrusted non-intrusively. The security risk is due to the design that the dynamic voltage and frequency of GPUs are not appropriately regulated and transient abnormal voltage-frequency glitches may induce hardware circuit timing constraint variations causing incorrect bit flips. A sensitive targets search algorithm and a genetic fault injection parameter search algorithm are proposed to address the major challenges in the attack target selection and the fault injection precision in terms of timing and position. Experiments on commodity Nvidia GPUs with four widely used CNNs verify the efficiency of the attack. Furthermore, the proposed attack can also support a high-performance targeted attack. As an innovative addition to the current DNN attacks from the perceptive of hardware vulnerability, this study reveals a potentially severe threat existing in hardware accelerator designs and also shows its notable impacts to the security of the DNN-related systems on shared GPU clouds. 

%In future work, more GPUs and models with larger data sets will be further evaluated, and different mitigating methods will also be realized and evaluated.

\balance
\bibliographystyle{ACM-Reference-Format}
\bibliography{reference}

\end{document}